\documentclass[twocolumn,aps,unsortedaddress]{revtex4-1}

\usepackage{amsmath}
\usepackage{mathtools}
\usepackage{bm}
\usepackage{amssymb}
\usepackage[colorlinks=true,linkcolor=blue,citecolor=blue,urlcolor=blue]{hyperref}
\usepackage{blkarray}

\newcommand{\Tr}[1]{\mathrm{Tr}{\left\{#1\right\}}}
\newcommand{\rhoss}{\hat{\rho}_{\mathrm{ss}}}
\newcommand{\Ropt}[1]{{#1}^{\mathrm{R}}}
\newcommand{\Aopt}[1]{{#1}^{\mathrm{A}}}
\newcommand{\RAopt}[1]{{#1}^{\mathrm{R/A}}}

\DeclarePairedDelimiter\abs{\lvert}{\rvert}
\DeclarePairedDelimiter\norm{\lVert}{\rVert}
\makeatletter
\let\oldabs\abs
\def\abs{\@ifstar{\oldabs}{\oldabs*}}
\let\oldnorm\norm
\def\norm{\@ifstar{\oldnorm}{\oldnorm*}}
\makeatother

\begin{document}

\title{Spin Polarization through A Molecular Junction Based on Nuclear Berry Curvature Effects}

\author{Hung-Hsuan Teh}
\email{teh@sas.upenn.edu}
\affiliation{Department of Chemistry, University of Pennsylvania, Philadelphia, Pennsylvania 19104, USA}

\author{Wenjie Dou}
\email{douwenjie@westlake.edu.cn}
\affiliation{Department of Chemistry, School of Science, Westlake University, Hangzhou, Zhejiang 310024, China}\textit{}
\affiliation{Department of Physics, School of Science, Westlake University, Hangzhou, Zhejiang 310024, China}\textit{}
\affiliation{Institute of Natural Sciences, Westlake Institute for Advanced Study, Hangzhou, Zhejiang 310024, China}


\author{Joseph E. Subotnik}
\email{subotnik@sas.upenn.edu}
\affiliation{Department of Chemistry, University of Pennsylvania, Philadelphia, Pennsylvania 19104, USA}

\date{\today}

\begin{abstract}
We explore the effects of spin-orbit coupling on nuclear wave packet motion near an out-of-equilibrium molecular junction, where nonzero Berry curvature emerges as the antisymmetric part of the electronic friction tensor.  The existence of nonzero Berry curvature mandates that different nuclear wave packets (associated with different electronic spin states) experience different nuclear Berry curvatures, i.e. different pseudo-magnetic fields. Furthermore, for a generic, two-orbital two-lead model (representing the simplest molecular junction), we report significant spin polarization of the {\em electronic} current with decaying and oscillating signatures in the large voltage limit --- all as a result of {\em nuclear} motion. These results are consistent with magnetic AFM chiral-induced spin selectivity experiments.  Altogether, our results  highlight an essential role for Berry curvature in condensed phase dynamics, where spin separation survives dissipation to electron-hole pair creation and emerges as one manifestation of nuclear Berry curvature.
\end{abstract}

\maketitle

Spintronics has a long history, originating with  the famous concept of giant magnetoresistance\cite{RevModPhys.80.1517}.
Nowadays, there are dozens of approaches for directly manipulating spin\cite{RevModPhys.76.323,RALPH20081190,doi:10.7566/JPSJ.82.102002,maekawa2017spin} including: various kinds of spin injection (e.g. based on transferring photonic angular momenta to electrons), spin pumping based on transferring spin angular momenta from magnetically precessing sources to conducting spin carriers, spin-transfer torque (which is the reverse of spin pumping), and spin Seebeck effects where a thermal gradient leads to a spin current in magnetic materials. The above-mentioned concepts can be realized not only in conventional inorganic materials but also for organic molecules\cite{C1CS15047B}.
Most recently, there has been an enormous amount of interest in the spin hall effect\cite{RevModPhys.82.1539} (which separates different spin carriers) and the spin quantum hall effect\cite{RevModPhys.82.3045} (which allows current to flow on the surface of topological insulators without back-scattering). There is no sign that progress in the arena of spintronics is slowing down.

Recently yet another approach for manipulating spin carriers has been proposed which is based on molecular chirality --- chiral-induced spin selectivity (CISS). Going back to early photoelectric experiments applied to DNA monolayers on metal surfaces\cite{doi:10.1126/science.1199339}, numerous experiments have since confirmed a CISS effect within various molecular monolayers\cite{kiran2017structure}, light-emitting diodes\cite{doi:10.1126/science.abf5291} and even single molecules\cite{aragones2017measuring}. In general, the CISS effect stipulates that, under finite voltage, the current running through a chiral set of molecules can be very spin-polarized. 
For our purposes below,  magnetic AFM CISS experiments have also been performed, whereby a monolayer of chiral molecules is in contact with a ferromagnetic material under an external magnetic field\cite{doi:10.1021/acs.jpclett.0c00474}; different majorities of spin sources are generated when  two opposite directions of the magnetic field are applied, and  correspondingly different currents are measured.

In order to explain the CISS effect, many explanations have been offered. Early on, scattering mechanisms were proposed, whereby a helical molecule can filter electrons according to angular momentum (which can be pinned to the SOC)\cite{gersten2013induced}. Dephasing approaches with leakage currents (and effectively multi-terminal physics) were also proposed\cite{PhysRevLett.108.218102}. More recently, investigations have been made  not only in the linear, but also in the nonlinear response regime\cite{PhysRevB.99.024418}.
Overall, there is today a reasonably large list of possible CISS mechanisms (e.g. \cite{doi:10.1021/jacs.1c05637}) --- though in  all cases, 
the necessary SOC required for large spin-polarization is still too large as compared with
\textit{ab initio} calculations\cite{maslyuk2018enhanced,doi:10.1021/acs.jctc.9b01078}. 
See Refs. \cite{PhysRevB.102.035445} and \cite{evers2021theory} for a summary of recent CISS theories.
At the moment, there is still no widely accepted theoretical understanding of the CISS effect.

Noting that nuclear motion is clearly important in DNA transport\cite{doi:10.1021/ja0540831,doi:10.1021/ar900123t,doi:10.1146/annurev-physchem-042018-052353}, recently we\cite{bian2021modeling} and Fransson\cite{PhysRevB.102.235416,fransson2021charge} have suggested that  CISS may arise from still another source: molecular vibrations. In particular, we have pointed out that CISS may well arise from semiclassical Berry forces.  Note that the influence of Berry forces is currently being explored within the quantum chemistry for small molecules with spin-orbit coupling\cite{culpitt2021ab}, but the effect of Berry forces in the condensed phase (with friction) is not well known. The goal of this letter is to explore if and how \textit{nuclear} dynamics and Berry force may in fact lead to the manipulation of \textit{electronic} spin in the condensed phase.
 
To explain our approach, consider  the standard Born-Oppenheimer (BO) approximation\cite{baer2006beyond}: a nuclear wave packet moves along on a surface corresponding to a certain electronic state $I$. The nuclear Hamiltonian is: $\mathbf{H}=(\mathbf{P}-\mathbf{A}^{I})^{2}/2M+\bm{\lambda}^{I}$ where $\mathbf{P}$ is the nuclear momentum, $\mathbf{A}^{I}\equiv i\hbar\langle I\vert\nabla\vert I\rangle$ is the nuclear Berry connection, and $\bm{\lambda}^{I}$ is the adiabatic surface of the state $\vert I\rangle$. The corresponding nuclear Berry curvature $\Omega_{ij}=\partial_{i}A_{j}^{I}-\partial_{j}A_{i}^{I}$ provides a pseudo-magnetic field in the nuclear $ij$-space. Notice that when a complex-valued Hamiltonian is considered, $\mathbf{A}^{I}$ does not vanish and the effect of $\Omega_{ij}$ must be included for accurate dynamics on surface $I$\cite{teh2021antisymmetric}.


Now, the discussion above was effectively predicated on modeling a small, isolated system with nuclear and electronic degrees of freedom. Within the condensed matter community,  the key question is whether such effects can be meaningful in the presence of dissipative channels. To that end, in order to treat a molecule on a metal surface (where there is a continuum of electronic bath states and the system is effectively open electronically),  one promising semiclassical approach is to apply an electronic friction tensor $\gamma_{\mu\nu}$ and random force $\zeta_{\mu}$ in addition to the adiabatic force $F_{\mu}$ where $\mu$ and $\nu$ label nuclear space directions\cite{bode2012current,smith1993electronic,PhysRevB.85.245444,PhysRevLett.119.046001,doi:10.1021/acs.jpca.8b09251}. According to such a treatment, which is valid both in and out of equilibrium\cite{bode2012current,doi:10.1021/nl904233u} provided that the electronic bath in the metal responds quickly and a Markovian ansatz is appropriate\cite{dou2018perspective}, a (molecular) nuclear degree of freedom near a metal surface is driven by a Langevin equation:
\begin{align}
M\ddot{R}_{\mu}=F_{\mu}-\sum_{\nu}\gamma_{\mu\nu}\dot{R}_{\nu}+\zeta_{\mu},\label{eq:langevin_eq}
\end{align}
Here, $M$ is the mass of nuclei and $R_{\mu}$ is the nuclear position in the $\mu$ direction.

When contemplating Eq. (\ref{eq:langevin_eq}), note that, at equilibrium, because of the fluctuation-dissipation theorem that relates $\gamma_{\mu \nu}$ to the covariance of the random force $\left<\zeta_{\mu} \zeta_{\nu} \right>$, one can use $F_{\mu}$ alone to determine the equilibrium density distribution ($\gamma_{\mu \nu}$ or $\left<\zeta_{\mu} \zeta_{\nu} \right>$ are not needed); see SM Sec. \ref{sec:fluctuation_dissipation} for a proof. Out of equilibrium, however,  there is no such fluctuation-dissipation theorem and there is indeed the possibility that the steady state distribution (and therefore steady state observables, e.g. the electronic current) {\em will} depend on $\gamma_{\mu \nu}$ and $\left<\zeta_{\mu} \zeta_{\nu} \right>$ as well as $F_{\mu}$.

At this point, let us address $\gamma_{\mu\nu}$.  The friction tensor $\gamma_{\mu\nu}$ in Eq. (\ref{eq:langevin_eq}) can be divided into a symmetric part $\gamma_{\mu\nu}^{\mathrm{S}}$ (which controls dissipative processes) and an antisymmetric part (which generates a Lorentz-like motion in the nuclear space and is effectively a generalization of the Berry curvature $\Omega_{ij}$).
The relative magnitudes of the symmetric and antisymmetric components can be calculated rigorously for a many-body Hamiltonian\cite{PhysRevLett.119.046001}. 
At equilibrium, previous work\cite{teh2021antisymmetric} has shown that the Lorentz-like force   $\gamma_{\mu\nu}^{\mathrm{A}}$ can be as large as  $\gamma_{\mu\nu}^{\mathrm{S}}$ (and sometimes even larger at low temperatures). In this letter, we will show that such a nuclear Lorentz-like force can actually lead to different electronic currents for different spin carriers in the presence of an nonequilibrium current and voltage for a model that roughly captures the diphenylmethane molecule in a junction.

Consider a simple model with two spatial orbitals (1 and 2) coupled to two leads, such that the Hamiltonian depends on two nuclear degrees of freedom ($x$ and $y$ are considered). The total  Hamiltonian $\hat{H}$ can be divided into four parts, the kinetic energy of nuclei, the subsystem (molecule) $\hat{H}_{\mathrm{s}}$, the bath continuum (two leads) $\hat{H}_{\mathrm{b}}$ and the subsystem-bath coupling $\hat{H}_{\mathrm{c}}$,
\begin{align}
&\hat{H}=\frac{P^{2}}{2M}+\hat{H}_{\mathrm{s}}+\hat{H}_{\mathrm{b}}+\hat{H}_{\mathrm{c}}\label{eq:total_e_H}\\
&\hat{H}_{\mathrm{s}}=\sum_{pq}h_{pq}^{\mathrm{s}}(\mathbf{R})\hat{b}_{p}^{\dagger}\hat{b}_{q}+U(\mathbf{R})\\
&\hat{H}_{\mathrm{b}}=\sum_{k,\alpha=\lbrace\mathrm{L},\mathrm{R}\rbrace}\epsilon_{k\alpha}\hat{c}_{k\alpha}^{\dagger}\hat{c}_{k\alpha}\\
&\hat{H}_{\mathrm{c}}=\sum_{p,k\alpha}V_{p,k\alpha}\hat{b}_{p}^{\dagger}\hat{c}_{k\alpha}+\mathrm{H.c.}\label{eq:Hc},
\end{align}
where $\hat{b}_{p}^{\dagger}$ ($\hat{b}_{p}$) creates (annihilates) an electron in the subsystem spin orbital $p$ (which is an element of  $\lbrace1,2\rbrace\otimes\lbrace\uparrow,\downarrow\rbrace$), $U(\mathrm{R})$ represents a nuclear-nuclear electrostatic potential (which does not depend on any electronic degree of freedom), $c_{k\alpha}^{\dagger}$ ($c_{k\alpha}$) creates (annihilates) an electron in the $k$-th spin orbital of the lead $\alpha$ (L/R denotes left/right lead) with the orbital energy $\epsilon_{k\alpha}$,  $V_{p,k\alpha}$ is the tunneling element between the subsystem spin orbital $p$ and the lead spin orbital $k\alpha$. In Eq. (\ref{eq:Hc}),  the  Condon approximation has been applied (so that $V_{p,k\alpha}$ does not depend on $\mathbf{R}$); typically speaking, one can roughly recover the size of the friction tensor without incorporating non-Condon effects\cite{dou2017electronic} .


In principle, all orbitals in Eqs. (\ref{eq:total_e_H})-(\ref{eq:Hc}) ($p$, $q$, $k$) are spin orbitals. However,  for a system with exactly two spatial orbitals, a key observation is that, at any fixed geometry, one can always define a new spin ``up'' basis $\vert\uparrow'\rangle$ which is  not coupled to the new spin ``down'' basis $\vert\downarrow'\rangle$. In other words,  we can postulate that there are \textit{two} independent Langevin equations driven by two friction tensors and random forces. This conclusion can be reached as follows (also see SM \ref{sec:block_diag} for details): A general spin-orbit interaction is $\xi\mathbf{L}\cdot\mathbf{S}$ where $\xi$ is the coupling strength, $\mathbf{L}$ is the angular momentum operator and $\mathbf{S}=\hbar\bm{\sigma}/2$ is the spin operator. Since $\mathbf{L}_{pp}$ vanishes in all spatial directions ($p=1,2$), the new spin basis can be obtained by diagonalizing $\xi\mathbf{L}_{12}\cdot\mathbf{S}$ which leads to eigenvalues $\pm a$ where $a=i\hbar\xi\abs{\mathbf{L}_{12}}/2$ is purely imaginary (note that $\mathbf{L}_{12}$ is purely imaginary). Thus, the subsystem Hamiltonian (at one geometry) becomes
\begin{align}\label{eq:hs_block_diag}
\mathbf{h}^{\mathrm{s}}=
\begin{pmatrix}
E_{1} & W & & \\
W & E_{2} & & \\
 & & E_{1} & W\\
 & & W & E_{2}
\end{pmatrix}
+\,\,\,\,
\begin{blockarray}{ccccc}
\begin{block}{(cccc)c}
0 & a & & & \,\,\vert1\uparrow'\rangle\\
a^{*} & 0 & & & \,\,\vert2\uparrow'\rangle\\
 & & 0 & -a & \,\,\vert1\downarrow'\rangle\\
 & & -a^{*} & 0 & \,\,\vert2\downarrow'\rangle\\
\end{block}
\end{blockarray},
\end{align}
where $E_{1,2}$ represent orbital energies and $W$ denotes the diabatic coupling between the two orbitals. For all of the discussion below, we will work in this new basis, and for notational simplicity we will discard the prime. Notice that the two $2\times2$ blocks, defined as $\mathbf{h}^{\mathrm{s}\uparrow}$ and $\mathbf{h}^{\mathrm{s}\downarrow}$, are complex conjugates of each other, and the pure spin rotation that diagonalizes the term $\xi\mathbf{L}_{12}\cdot\mathbf{S}$ does not affect the orbital energies and couplings.  According to Eq. (\ref{eq:hs_block_diag}), if we  assume that the spin basis does not change (or changes weakly as a function of $\mathbf{R}$), then there is no interaction between two spin carriers, and we may propagate the two spin degrees of freedom separately (with $\mathbf{h}^{\mathrm{s}\uparrow}$ and $\mathbf{h}^{\mathrm{s}\downarrow}$). Obviously, if there is no spin-orbit interaction, and $\mathbf{h}^{\mathrm{s}\uparrow}=\mathbf{h}^{\mathrm{s}\downarrow}$, the dynamics of the different spin carriers will be the same. However, when $\mathbf{h}^{\mathrm{s}\uparrow} \ne \mathbf{h}^{\mathrm{s}\downarrow}$, we will show that substantially different spin currents can arise.

\begin{figure}
\centering
\includegraphics[width=0.48\textwidth]{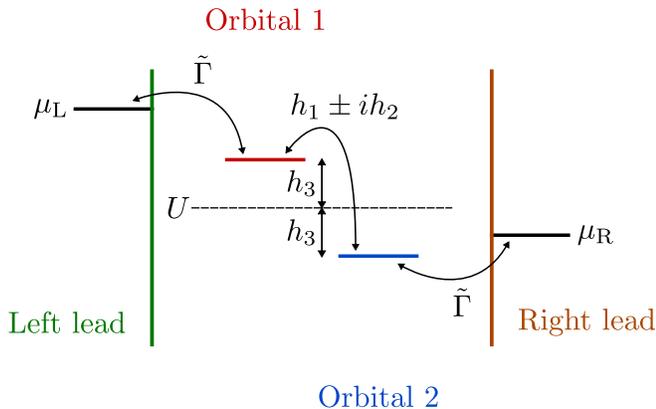}
\caption{A schematic picture of our molecular junction model. $\mu_{\mathrm{L}}$ and $\mu_{\mathrm{R}}$ are chemical potentials of the left and right leads, and the source-drain voltage  $V_{sd}=\mu_{\mathrm{L}}-\mu_{\mathrm{R}}$ controls the voltage bias. Other parameters are defined in the text.}\label{fig:schematic_pic}
\end{figure}

\begin{figure}
\centering
\includegraphics[width=0.48\textwidth]{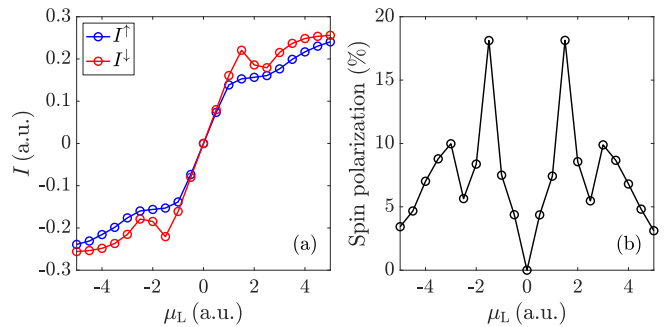}
\caption{$\Delta=0$, $A=B=1$, $\kappa_{x}=0$, $\chi=1$, $\kappa_{y}=0.8$, $\tilde{\Gamma}=1$ and $\mu_{\mathrm{L}}=-\mu_{\mathrm{R}}$. We utilize Eq. (\ref{eq:spin_current}) to calculate (a) the spin current and (b) the corresponding spin polarization. No spin polarization is predicted when the voltage bias is zero. Sizable spin polarization can be found at finite $\mu_{\mathrm{L}}$. The decaying and the oscillating behaviors at large bias limit are consistent with an AFM CISS experiment.}\label{fig:spin_current_n_polarization}
\end{figure}

Consider the common case where orbital $1$ couples only to the left lead, and orbital $2$ couples only to the right lead (see Fig. \ref{fig:schematic_pic}).  For the Hamiltonian in Eq. (\ref{eq:hs_block_diag}),
let $\mu_L$ and $\mu_R$ be the fermi levels in the left and right leads, with voltage $V \equiv \mu_L - \mu_R$ across the molecule.  We
run Langevin dynamics and  calculate the spin current $I^{\uparrow/\downarrow}$ by taking an  ensemble average over nuclear DoFs,
\begin{align}
I^{\uparrow/\downarrow}=\int d\mathbf{R}d\mathbf{P}\,I_{\mathrm{loc}}^{\uparrow/\downarrow}(\mathbf{R})\rho^{\uparrow/\downarrow}(\mathbf{R},\mathbf{P}).\label{eq:spin_current}
\end{align}
The nuclear probability density $\rho^{\uparrow/\downarrow}(\mathbf{R},\mathbf{P})$ is determined by sampling 1000 trajectories, running dynamics according to Eq. (\ref{eq:langevin_eq}), and evaluating how many trajectories are at ($\mathbf{R}$,$\mathbf{P}$) in phase space at steady state. General properties of the friction tensor formalism and all relevant matrix elements are shown in the SM Secs. \ref{sec:sum_friction_tensor}-\ref{sec:adiaF}.  We make the ansatz that the local spin current $I_{\mathrm{loc}}^{\uparrow/\downarrow}$ flowing from the left lead through the molecule to the right lead can be evaluated by the Landauer formula\cite{haug2008quantum},
\begin{align}\label{eq:Landauer_formula_local_current}
I_{\mathrm{loc}}^{\uparrow/\downarrow}=\frac{e}{2\pi\hbar}\int_{-\infty}^{\infty}d\epsilon\,T^{\uparrow/\downarrow}(\epsilon)\left[f_{\mathrm{L}}(\epsilon)-f_{\mathrm{R}}(\epsilon)\right],
\end{align}
where $T^{\uparrow/\downarrow}(\epsilon)$ is the transmission probability that depends on $h^{2}$ (see SM \ref{sec:transmission_prob} for details).  Note that one cannot distinguish the two spin carriers if the two probability densities $\rho^{\uparrow/\downarrow}$ are the same. Note also that, for the case of  a single resonant level, the current calculated by Eq. (\ref{eq:spin_current}) has been shown to agree with numerically exact HEOM calculations\cite{dou2018broadened}.

We focus on the  shifted parabola model that is elaborated on in SM Sec. \ref{sec:shifted_parabola} as a rough model for a diphenylmethane molecule in a  junction:
\begin{align}
&\mathbf{h}^{\mathrm{s}\uparrow}=
\begin{pmatrix}
x+\Delta & Ax-iBy\\
Ax+iBy & -x-\Delta
\end{pmatrix},\label{eq:hs_spin_up}\\
&U(\mathbf{R})=\frac{1}{2}x^{2}+\kappa_{x}x+\frac{1}{2}\chi y^{2}+\kappa_{y}y.\label{eq:U}
\end{align}
Here $\Delta$ tunes the energy gap between two orbitals, $A$ and $B$ control the rates at which the diabatic and spin-orbit couplings change (respectively) as a function of geometries $x$ and $y$. We include a scalar potential $U(\mathbf{R})$
that tilts the overall energy landscape; such a tilt has been shown to be crucial when understanding the dynamics of photoexcited molecules relaxing through conical intersections\cite{malhado2012photoisomerization}.  Note that this potential does not affect the electronic friction in any way. The linear terms $\kappa_{x}x$ and $\kappa_{y}y$ 
effectively decrease the symmetry of the total adiabatic state and 
$\chi$ is the ratio of the mode-$y$ frequency to the mode-$x$ frequency.  In practice, the scalar terms have nothing to do with spin, but as discussed below, these terms can be crucial for generating a strong spin current.

In Fig. \ref{fig:spin_current_n_polarization} we plot the spin current (calculated by utilizing Eq. (\ref{eq:Landauer_formula_local_current})) and the corresponding spin polarization results  in the symmetric case ($\Delta=0$) with  $\mu_{\mathrm{L}}=-\mu_{\mathrm{R}}$. The spin polarization is defined as the standard quantity $(I^{\downarrow}-I^{\uparrow})/(I^{\downarrow}+I^{\uparrow})$. For this initial set of data, we simulate a large SOC; we set $A=B=\chi=1$ so that the average $\left<Ax\right>$ is of the same order of magnitude as $\left<By\right>$.  We find an $18\%$ spin polarization; furthermore we find decaying and the oscillating behaviors in the large bias limit, which is consistent with the magnetic AFM results in Ref. \citenum{doi:10.1021/acs.jpclett.0c00474}. We emphasize that in SM \ref{sec:small_soc}, we consider the case where there is a much smaller spin-orbit interaction ($B=0.1$ and  $\chi=0.1$ so that $\left<Ax\right> \ll \left<By\right>$); a sizable spin polarization is still achieved at large voltages.



\begin{figure}
\centering
\includegraphics[width=.48\textwidth]{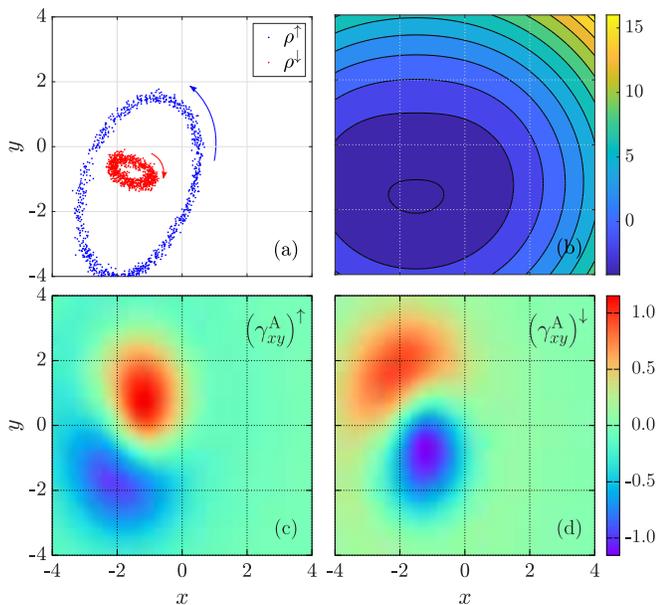}
\caption{(a) Scatter plot distribution of spin up/down carriers in steady-state with an overall arrow indicating the direction of motion. (b) Adiabatic ground state obtained from diagonalizing $\mathbf{h}^{\mathrm{s}\uparrow}+U$. (c) Antisymmetric friction tensor for $B=1$ (spin up). (d) Antisymmetric friction tensor for $B=-1$ (spin down). The combination of the adiabatic force and the different pseudo-magnetic fields for spin up and down leads to different steady-state nuclear probability distributions.}\label{fig:muL3}
\end{figure}

To date, we have run several calculations for symmetric systems ($\Delta = 0$) and asymmetric systems ($\Delta \ne 0$). Overall,  our results show a surprising richness of spin-dependent phenomena, including the fact that setting $\Delta \ne 0$ tends to  enhance the spin polarization (see SM \ref{sec:D3} for details). In general, we find that spin current results depend sensitively on the global nature of the potential energy surface (and not just the spin-orbit coupling). 

In order to illustrate microscopic origins of spin polarization, we consider an asymmetric Hamiltonian (with $\Delta=3$, $\kappa_{x}=0$, and $\kappa_{y}=1$). The I-V curve can be found in the SM \ref{sec:D3}.  For such a Hamiltonian, a large spin-current is found when $\mu_{L}= -\mu_R=3.5$. In Fig. \ref{fig:muL3} (c) and (d), we plot the antisymmetric friction tensors (corresponding to the pseudo-magnetic field) when $B=1$ and $B=-1$ respectively. The adiabatic ground state potential is plotted in Fig. \ref{fig:muL3} (b).  For such a Hamiltonian, one finds  very well separated nuclear nonequilibrium steady state distributions in position space, $\rho^{\uparrow/\downarrow}(\mathbf{R},\mathbf{P})$, for different spin carriers as shown in Fig. \ref{fig:muL3} (a) and this separation of nuclear densities leads to different spin currents. To explain why the nuclear steady state densities are so different {\em out of equilibrium} (even though they must agree at equilibrium), note that the spin up carriers are affected both by the two opposite pseudo-magnetic fields (blue and red regions in Fig. \ref{fig:muL3} (c)) and the (adiabatic) restoring force in Fig. \ref{fig:muL3} (b). As  a result, the trajectories trace out the big circulating vector field (with an overall blue arrow indicating the direction of motion) in Fig. \ref{fig:muL3} (a). By contrast, the spin down carriers experience only one kind of pseudo-magnetic field (the blue region in Fig. \ref{fig:muL3} (d)) so that their steady state nuclear distribution does not stray far from the equilibrium region (see the red dots in Fig. \ref{fig:muL3} (a)). This nonequilibrium difference in steady state nuclear distributions leads to clear differences in spin currents. As a side note, the nuclear Berry curvature effect becomes small when the voltage is very large and, in such a case, we predict that a distinction between spins can no longer be achieved. This prediction is in agreement with the AFM-CISS experiment, whereby the difference in current (between systems with up and down magnetic fields) is found to decrease at large voltages.\cite{doi:10.1021/acs.jpclett.0c00474}

Looking forward, there are many questions that must be addressed. First and most importantly, the model presented here is clearly just a model: in the future, one would like to run fully {\em ab initio} dynamics without any parameters.  
Second, the strong dependence of the spin current on the $U$ term in Eq. (\ref{eq:U}) is interesting and highlights the fact that, if electronic spin transitions are coupled to nuclear dynamics, then understanding spin dynamics may necessarily require modeling the full totality of chemical dynamics; spin-orbit coupling will not be the only determinant of spin-current. In fact, optimizing spin polarization in practice (experimentally) might require optimizing nuclear barriers (rather than just increasing SOC).  Third, in this letter, we have focused explicitly on systems with only two orbitals, and for such a system, one cannot extract a spin-current without nuclear motion.  However, if we allow for three or more orbitals (and freeze the nuclei), one can extract different (but small) spin currents based on purely electronic considerations. How will the spin polarization based on nuclear nonadiabatic motion scale with system size, especially if we were to treat a true helix?  Note that CISS experiments on DNA are very sensitive to the length of the DNA\cite{doi:10.1021/jz300793y}. There are many exciting questions to answer in the future.

In conclusion, we have demonstrated that, in the presence of a nonzero electronic current,  Berry curvature effects can lead to spin separation of nuclear wave packets and spin-polarization of the electronic current.  These effects arise because  $(i)$ the presence of  spin-orbit coupling creates a Berry force that does not obey time-reversal symmetry within a master equation model, and $(ii)$ the presence of a finite voltage leads to a random force that does not obey a fluctuation-dissipation theorem. Altogether, these two effects lead to the scenario whereby the steady states for nuclear wave packets with spin up electrons can be very different from the steady states for nuclear wave packets with spin down electrons. 
Importantly, we find that a sizable spin polarization can be achieved even in the situation where the spin-orbit coupling is small (a tenth of a diabatic coupling) when modes with different frequencies are considered; future research will be necessary to address exactly how small the spin-orbit coupling can be and still yield a meaningful effect --- especially given the possibility of more than two orbitals and possible conical intersections.
(Note that our model has made the simplifying approximation that different spin carriers do not mix with each other, 
and future work will also be necessary to explore the implication of spin-flip processes.)
Lastly,  our results in Fig. \ref{fig:muL3} have shown that, as a result of nuclear motion, decaying and oscillating signatures in the polarization can emerge as a function of voltage for the magnetic AFM setup (which has been observed experimentally \cite{doi:10.1021/acs.jpclett.0c00474}).  Altogether, the present letter suggests that a merger of spintronics and nonadiabatic dynamics is on the horizon, and the experimental observation of nonequilibrium spin separation and polarization in chemical systems would appear to be the glue that connects together these two titanic fields of condensed matter physics.


{\bf Acknowledgment}
Hung-Hsuan Teh thanks Yi-Hsien Liu and Sijia Ke for helpful discussions. This work was supported by the U.S. Air Force Office of Scientific Research (USAFOSR) under Grants No. FA9550- 18-1-0497 and No. FA9550-18-1-0420.

\bibliography{main}

\onecolumngrid
\newpage

\renewcommand{\thesection}{\Alph{section}}
\renewcommand{\theequation}{\thesection\arabic{equation}}
\renewcommand{\thefigure}{\thesection\arabic{figure}}
\setcounter{section}{0}
\setcounter{equation}{0}
\setcounter{figure}{0}
\setcounter{page}{1}

\begin{center}
\bf\large Supplemental Material --- Spin Polarization through A Molecular Junction Based on Nuclear Berry Curvature Effects
\end{center}

\begin{center}
Hung-Hsuan Teh\\
\textit{Department of Chemistry, University of Pennsylvania, Philadelphia, Pennsylvania 19104, USA}
\end{center}

\begin{center}
Wenjie Dou\\
\textit{Department of Chemistry, School of Science, Westlake University, Hangzhou, Zhejiang 310024, China}\\
\textit{Department of Physics, School of Science, Westlake University, Hangzhou, Zhejiang 310024, China}\\
\textit{Institute of Natural Sciences, Westlake Institute for Advanced Study, Hangzhou, Zhejiang 310024, China}
\end{center}

\begin{center}
Joseph E. Subotnik\\
\textit{Department of Chemistry, University of Pennsylvania, Philadelphia, Pennsylvania 19104, USA}
\end{center}

\section{Summary of the Friction Tensor}\label{sec:sum_friction_tensor}
In this section, we will briefly review the friction tensor $\gamma_{\mu\nu}$ in Eq. (\ref{eq:langevin_eq}) in the main body of the text. The equation of motion driving the nuclear probability density $\rho(\mathbf{R},\mathbf{P})$ (for notational simplicity here we consider the spinless case) can be derived from the mixed quantum-classical Liouville equation\cite{doi:10.1146/annurev.physchem.57.032905.104702} followed by the Mori-Zwanzig method and the adiabatic approximation\cite{PhysRevLett.119.046001},
\begin{align}
\partial_{t}\rho=-\sum_{\mu}\frac{P_{\mu}}{m_{\mu}}\partial_{\mu}\rho-\sum_{\mu}F_{\mu}\frac{\partial\rho}{\partial P_{\mu}}+\sum_{\mu\nu}\gamma_{\mu\nu}\frac{\partial}{\partial P_{\mu}}\left(\frac{P_{\nu}}{m_{\nu}}\rho\right)+\sum_{\mu\nu}\bar{D}_{\mu\nu}^{\mathrm{S}}\frac{\partial^{2}\rho}{\partial P_{\mu}\partial P_{\nu}},\label{eq:fokker_planck}
\end{align}
where in this Fokker-Planck equation, which is equivalent to Eq. (\ref{eq:langevin_eq}), the adiabatic force $F_{\mu}$, friction tensor $\gamma_{\mu\nu}$ and covariance matrix  $\bar{D}_{\mu\nu}^{\mathrm{S}}$ for the random force $\zeta_{\mu}$ (in Eq. (\ref{eq:langevin_eq})) are
\begin{align}
F_{\mu}&=-\Tr{\partial_{\mu}\hat{H}\rhoss},\label{eq:adia_F}\\
\gamma_{\mu\nu}&=-\int_{0}^{\infty}dt\,\Tr{\partial_{\mu}\hat{H} e^{-i\hat{H}t/\hbar} \partial_{\nu}\rhoss e^{i\hat{H}t/\hbar}},\\
\bar{D}_{\mu\nu}^{\mathrm{S}}&=\frac{1}{2}\int_{0}^{\infty}dt\,\Tr{e^{i\hat{H}t/\hbar} \delta\hat{F}_{\mu} e^{-i\hat{H}t/\hbar} \left(\delta\hat{F}_{\nu}\rhoss+\rhoss\delta\hat{F}_{\nu}\right)},\label{eq:D}\\
\delta\hat{F}_{\mu}&=-\partial_{\mu}\hat{H}+\Tr{\partial_{\mu}\hat{H}\rhoss}.\label{eq:deltaF}
\end{align}
Here $\hat{H}$ is the electronic Hamiltonian, $\rhoss$ is the steady state density matrix satisfying $[\hat{H},\rhoss]=0$, and $\bar{D}_{\mu\nu}^{\mathrm{S}}$ is in the Markovian limit such that the random force $\zeta_{\mu}(t)$ satisfies the time correlation function,
\begin{align*}
\frac{1}{2}\left[\langle\zeta_{\mu}(t)\zeta_{\nu}(t')\rangle+\langle\zeta_{\nu}(t)\zeta_{\mu}(t')\rangle\right]=\bar{D}_{\mu\nu}^{\mathrm{S}}\delta(t-t').
\end{align*}

When a non-interacting Hamiltonian $\hat{H}=\sum_{pq}\mathcal{H}_{pq}(\mathbf{R})\hat{d}_{p}^{\dagger}\hat{d}_{q}+U(\mathbf{R})$ is considered ($\hat{d}_{p}^{\dagger}$/$\hat{d}_{p}$ creates/annihilates an electron in orbital $p$, and $U(\mathbf{R})$ is a purely nuclear potential energy), the friction tensor becomes\cite{PhysRevB.97.064303}
\begin{align*}
\gamma_{\mu\nu}=-\frac{1}{2\pi}\int_{-\infty}^{\infty}d\epsilon\,\Tr{\partial_{\mu}\mathcal{H}\Ropt{\mathcal{G}}\partial_{\nu}\sigma_{\mathrm{ss}}\Aopt{\mathcal{G}}},
\end{align*}
where $\RAopt{\mathcal{G}}=(\epsilon-\mathcal{H}\pm i\eta)^{-1}$ is the retarded/advanced Green's function of the electron, and
\begin{align}
\sigma_{qp}^{\mathrm{ss}}\equiv\Tr{\rhoss\hat{d}_{p}^{\dagger}\hat{d}_{q}}=\int_{-\infty}^{\infty}\frac{d\epsilon}{2\pi i}\mathcal{G}_{qp}^{<}(\epsilon).\label{eq:glesser_n_sigma}
\end{align}
Here $\mathcal{G}_{qp}^{<}(\epsilon)$ is the lesser Green's function in the energy domain. Here we have used the fact that $\mathcal{G}_{qp}^{<}(t_{1},t_{2})=\mathcal{G}_{qp}^{<}(t_{2}-t_{1})$, due to $[\hat{H},\rhoss]=0$, so that the conventional time-domain lesser Green's function,
\begin{align*}
\mathcal{G}_{qp}^{<}(t_{1},t_{2})\equiv\frac{i}{\hbar}\Tr{\rhoss\hat{d}_{p}^{\dagger}(t_{2})\hat{d}_{q}(t_{1})},
\end{align*}
can be Fourier transformed.

In order to proceed, $\mathcal{G}^{<}(\epsilon)$ is constructed to follow the Keldysh equation $\mathcal{G}^{<}=\Ropt{\mathcal{G}}\Pi^{<}\Aopt{\mathcal{G}}$ (this is true when the relaxation from the system described by $\hat{H}$ to a fictitious outer bath is fast enough\cite{PhysRevB.97.064303}) where $\Pi^{<}$ is the electron lesser self energy assumed to be independent of $\epsilon$. Then the friction tensor $\gamma_{\mu\nu}$ becomes (Ref. \cite{PhysRevB.97.064303})
\begin{align}
\gamma_{\mu\nu}=\frac{\hbar}{2\pi}\int_{-\infty}^{\infty}d\epsilon\,\Tr{\partial_{\mu}\mathcal{H}\partial_{\epsilon}\Ropt{\mathcal{G}}\partial_{\nu}\mathcal{H}\mathcal{G}^{<}}+\mathrm{H.c.}.\label{eq:gamma_noninteracting_antiH_glesser}
\end{align}
In equilibrium, as shown in Ref. \citenum{teh2021antisymmetric}, the antisymmetric part of Eq. (\ref{eq:gamma_noninteracting_antiH_glesser}) can be simplified. The result is:
\begin{align}
\gamma_{\mu\nu}^{\mathrm{A}}\propto-\sum_{k\neq l,\epsilon_{k}\neq\epsilon_{l}}2\mathfrak{Im}\left\{d_{kl}^{\mu}d_{lk}^{\nu}\right\}\left[f(\epsilon_{k})-f(\epsilon_{l})\right],\label{eq:gammaA_dc_n_fermi}
\end{align}
where $d_{kl}^{\mu}\equiv\langle k\vert\partial_{\mu}\vert l\rangle$ is the derivative coupling between Lehmann representations, and $f(\epsilon)=1/[\exp{(\beta(\epsilon-\mu))}+1]$ represents Fermi-Dirac distribution. We emphasize that Eq. \ref{eq:gammaA_dc_n_fermi} is valid only in equilibrium.

At this point,   we consider the molecular junction Hamiltonian (Eqs. (\ref{eq:total_e_H})-(\ref{eq:Hc}) in the main body of the text). We will then replace $\mathcal{H}$ with $h$ (and $\mathcal{G}$ with G). 
For an arbitrary voltage, Eq. (\ref{eq:gamma_noninteracting_antiH_glesser}) can be further simplified under the Condon approximation where the coupling $V_{p,k\alpha}$ is independent of the nuclear position $\mathbf{R}$. An analytic expression for the friction tensor for a general two-orbital two-mode Hamiltonian (namely $\mathbf{h}^{\mathrm{s}\uparrow}=\mathbf{h}(x,y)\cdot\bm{\sigma}$) was derived in Ref. \cite{teh2021antisymmetric}, and the result is
\begin{alignat}{2}
\gamma_{\mu\nu}=&\gamma_{\mu\nu}^{\mathrm{S}}+\gamma_{\mu\nu}^{\mathrm{A}},\label{eq:ft_tls}\\
\gamma_{\mu\nu}^{\mathrm{S}}=&\frac{2}{\pi}\int_{-\infty}^{\infty}d\epsilon\bigg\{&&-2\mathfrak{Re}\left\{C\tilde{\epsilon}\right\}\left(\partial_{\mu}\mathbf{h}\cdot\partial_{\nu}\mathbf{h}\right)\left(\mathbf{h}\cdot\bm{\kappa}\right)\notag\\
&&&+2\mathfrak{Re}\left\{C\tilde{\epsilon}\right\}\left(\partial_{\mu}\mathbf{h}\cdot\mathbf{h}\right)\left(\partial_{\nu}\mathbf{h}\cdot\bm{\kappa}\right)\notag\\
&&&+2\mathfrak{Re}\left\{C\tilde{\epsilon}\right\}\left(\partial_{\nu}\mathbf{h}\cdot\mathbf{h}\right)\left(\partial_{\mu}\mathbf{h}\cdot\bm{\kappa}\right)\notag\\
&&&+\kappa_{0}\mathfrak{Re}\left\{C\left(\tilde{\epsilon}^{2}+h^{2}\right)\right\}\partial_{\mu}\mathbf{h}\cdot\partial_{\nu}\mathbf{h}\bigg\}\label{eq:ft_tls_sym}\\
\gamma_{\mu\nu}^{\mathrm{A}}=&\frac{2}{\pi}\int_{-\infty}^{\infty}d\epsilon\bigg\{&&-\mathfrak{Im}\left\{C\left(\tilde{\epsilon}^{2}+h^{2}\right)\right\}
\bm{\kappa}\cdot\left(\partial_{\mu}\mathbf{h}\times\partial_{\nu}\mathbf{h}\right)\notag\\
&&&+2\kappa_{0}\mathfrak{Im}\left\{C\tilde{\epsilon}\right\}\mathbf{h}\cdot\left(\partial_{\mu}\mathbf{h}\times\partial_{\nu}\mathbf{h}\right)\bigg\},\label{eq:ft_tls_antisym}
\end{alignat}
where $C\equiv-\left(\frac{1}{\tilde{\epsilon}^{2}-h^{2}}\right)^{2}i\tilde{\Gamma}\abs{\frac{1}{\tilde{\epsilon}^{2}-h^{2}}}^{2}$ and $\tilde{\epsilon}=\epsilon+i\tilde{\Gamma}/2$ is a complex number. $\tilde{\Gamma}$ represents the molecule-lead coupling strength, which is a constant under the wide-band-limit approximation, i.e.
$\tilde{\Gamma} = \Gamma_{11} = \Gamma_{22}$.  Here, we have defined
 $\Gamma_{mn}\equiv2\pi\sum_{k\alpha}V_{m,k\alpha}V_{n,k\alpha}^{*}\delta(\epsilon-\epsilon_{k\alpha})$;
 note that the off-diagonal elements $\Gamma_{12}$ and $\Gamma_{21}$ are zero because orbital $1$ couples only to the left lead and orbital $2$ couples only to the right lead. 
 
 The components of $\bm{\kappa}$ are
\begin{align*}
\kappa_{0}=&\frac{1}{2}\left[\left(f_{\mathrm{L}}+f_{\mathrm{R}}\right)\left(h_{1}^{2}+h_{2}^{2}\right)+f_{\mathrm{L}}\abs{\tilde{\epsilon}+h_{3}}^{2}+f_{\mathrm{R}}\abs{\tilde{\epsilon}-h_{3}}^{2}\right],\\
\kappa_{1}=&\mathfrak{Re}\left\{\left[f_{\mathrm{L}}\left(\tilde{\epsilon}^{*}+h_{3}\right)+f_{\mathrm{R}}\left(\tilde{\epsilon}-h_{3}\right)\right]\left(h_{1}+ih_{2}\right)\right\},\\
\kappa_{2}=&\mathfrak{Im}\left\{\left[f_{\mathrm{L}}\left(\tilde{\epsilon}^{*}+h_{3}\right)+f_{\mathrm{R}}\left(\tilde{\epsilon}-h_{3}\right)\right]\left(h_{1}+ih_{2}\right)\right\},\\
\kappa_{3}=&\frac{1}{2}\left[\left(f_{\mathrm{R}}-f_{\mathrm{L}}\right)\left(h_{1}^{2}+h_{2}^{2}\right)+f_{\mathrm{L}}\abs{\tilde{\epsilon}+h_{3}}^{2}-f_{\mathrm{R}}\abs{\tilde{\epsilon}-h_{3}}^{2}\right].
\end{align*}
Notice that all of the $\kappa$'s are real functions. Also, when the total system is in equilibrium, i.e. when $f_{L}=f_{R}=f$,
\begin{align*}
\kappa_{0}=&f\left(\epsilon^{2}+h^{2}+\frac{\Gamma^{2}}{4}\right),\\
\bm{\kappa}=&2f\epsilon\mathbf{h}.
\end{align*}
Equations (\ref{eq:ft_tls})-(\ref{eq:ft_tls_antisym}) are used in propagating Eq. (\ref{eq:langevin_eq}).

\section{Expression of $\bar{D}_{\mu\nu}^{\mathrm{S}}$ in Terms of Green's Functions}\label{sec:D_in_terms_of_G}
In this section, we will derive a practical expression of the covariance matrix $\bar{D}_{\mu\nu}^{\mathrm{S}}$ in terms of Green's functions. All of the approximations invoked below are consistent with the derivation of the electronic friction tensor in Eqs. (\ref{eq:ft_tls})-(\ref{eq:ft_tls_antisym}) (see Ref. \cite{teh2021antisymmetric} for details). 

To proceed, we first note that when a non-interacting Hamiltonian is considered, $U(\mathbf{R})$ does not contribute to $\delta\hat{F}_{\mu}$, since according to Eq. (\ref{eq:deltaF}),
\begin{align}
\delta\hat{F}_{\mu}=-\sum_{pq}\partial_{\mu}\mathcal{H}_{pq}\left(\hat{d}_{p}^{\dagger}\hat{d}_{q}-\sigma_{qp}^{\mathrm{ss}}\right).\label{eq:deltaF_noninteracting}
\end{align}
Furthermore, since $U(\mathbf{R})$ is a scalar function, it does not contribute to $\bar{D}_{\mu\nu}^{\mathrm{S}}$ according to Eq. (\ref{eq:D}). Second, according to Eq. (\ref{eq:D}), $\bar{D}_{\mu\nu}^{\mathrm{S}}$ consists of two parts involving $\delta\hat{F}_{\nu}\rhoss$ and $\rhoss\delta\hat{F}_{\nu}$ respectively, and the two parts are Hermitian conjugate to each other. We focus on the former and substitute Eq. (\ref{eq:deltaF_noninteracting}) for $\delta\hat{F}_{\mu}$ in Eq. (\ref{eq:D}):
\begin{align*}
&\frac{1}{2}\int_{0}^{\infty}dt\,\Tr{e^{i\hat{H}t/\hbar} \delta\hat{F}_{\mu} e^{-i\hat{H}t/\hbar} \delta\hat{F}_{\nu}\rhoss}\\
=&\frac{1}{2}\int_{0}^{\infty}dt\,\Tr{e^{i\hat{H}t/\hbar}
\sum_{pq}\partial_{\mu}\mathcal{H}_{pq}\left(\hat{d}_{p}^{\dagger}\hat{d}_{q}-\sigma_{qp}^{\mathrm{ss}}\right)
e^{-i\hat{H}t/\hbar}
\sum_{rs}\partial_{\nu}\mathcal{H}_{rs}\left(\hat{d}_{r}^{\dagger}\hat{d}_{s}-\sigma_{sr}^{\mathrm{ss}}\right)
\rhoss}\notag\\
=&\frac{1}{2}\int_{0}^{\infty}dt\,\Tr{\partial_{\mu}\mathcal{H} e^{-i\mathcal{H}t/\hbar} (1-\sigma^{\mathrm{ss}}) \partial_{\nu}\mathcal{H} \sigma^{\mathrm{ss}} e^{i\mathcal{H}t/\hbar}}.
\end{align*}
Here, we have utilized Wick's theorem to evaluate a two particle Green's function $\Tr{\hat{d}_{a}^{\dagger}\hat{d}_{b}\hat{d}_{r}^{\dagger}\hat{d}_{s}\rhoss}$ (see Eqs. (5.1) and (5.27) in Ref. \cite{stefanucci2013nonequilibrium}),
\begin{align*}
\Tr{\rhoss\hat{d}_{a}^{\dagger}(4)\hat{d}_{b}(3)\hat{d}_{r}^{\dagger}(2)\hat{d}_{s}(1)}
&=-\Tr{\mathcal{T}\left[\rhoss\hat{d}_{b}(3)\hat{d}_{s}(1)\hat{d}_{a}^{\dagger}(4)\hat{d}_{r}^{\dagger}(2)\right]}\\
&=G_{2}(3,1;2,4)\\
&=\sigma_{sa}^{\mathrm{ss}}\left(\delta_{br}-\sigma_{br}^{\mathrm{ss}}\right)+\sigma_{ba}^{\mathrm{ss}}\sigma_{sr}^{\mathrm{ss}}.
\end{align*}

We proceed to write Eq. (\ref{eq:D}) in the energy domain,
\begin{align}
\bar{D}_{\mu\nu}^{\mathrm{S}}=\frac{\hbar}{4\pi}
\int_{-\infty}^{\infty}d\epsilon\,\Tr{\partial_{\mu}\mathcal{H} \frac{1}{\epsilon-\mathcal{H}+i\eta} (1-\sigma^{\mathrm{ss}}) \partial_{\nu}\mathcal{H} \sigma^{\mathrm{ss}} \frac{1}{\epsilon-\mathcal{H}-\i\eta}}+\mathrm{H.c.},\label{eq:D_energy_domain}
\end{align}
where we have used integral representations of the Dirac delta function and the Heaviside function.

In order to evaluate $\bar{D}_{\mu\nu}^{\mathrm{S}}$ in practice, we hope to express Eq. (\ref{eq:D_energy_domain}) in terms of Green's functions. We expand Eq. (\ref{eq:D_energy_domain}) in an orbital basis and utilize the residue theorem to evaluate the integral over $\epsilon$, obtaining
\begin{align}
\bar{D}_{\mu\nu}^{\mathrm{S}}=i\frac{\hbar}{2}
\sum_{pqrs}(\partial_{\mu}\mathcal{H})_{pq} \frac{1}{\epsilon_{p}-\epsilon_{q}+i\eta} (1-\sigma^{\mathrm{ss}})_{qr} (\partial_{\nu}\mathcal{H})_{rs} \sigma_{sp}^{\mathrm{ss}}
+\mathrm{H.c.}\label{eq:D_energy_domain_orb_basis}
\end{align}
Then we replace $\sigma^{\mathrm{ss}}$ by using Eq. (\ref{eq:glesser_n_sigma}). Next, we further assume that the relaxation from the system modeled by $\hat{H}$ (more specifically from the bath Hamiltonian $\hat{H}_{\mathrm{b}}$) as caused by a fictitious outer bath is fast enough so that we can utilize the Keldysh relation,
\begin{align}
\mathcal{G}^{<}(\epsilon)=\Ropt{\mathcal{G}}(\epsilon)\Pi^{<}\Aopt{\mathcal{G}}(\epsilon),\label{eq:keldysh}
\end{align}
where the lesser self-energy is again assumed to be independent of $\epsilon$. As a result, Eq. (\ref{eq:glesser_n_sigma}) becomes
\begin{align*}
\sigma_{qr}^{\mathrm{ss}}\simeq\frac{1}{2\pi i}\int_{-\infty}^{\infty}d\epsilon\,\left(\Ropt{\mathcal{G}}(\epsilon)\Pi^{<}\Aopt{\mathcal{G}}(\epsilon)\right)_{qr}
=\frac{1}{\epsilon_{r}-\epsilon_{q}+i\eta}\Pi^{<}_{qr}.
\end{align*}

Note that there are two contributions in Eq. (\ref{eq:D_energy_domain_orb_basis}): one with $\delta_{qr}$ and the other with $\sigma_{qr}^{\mathrm{ss}}$. We first address the former,
\begin{align}
&i\frac{\hbar}{2}
\sum_{pqrs}(\partial_{\mu}\mathcal{H})_{pq} \frac{1}{\epsilon_{p}-\epsilon_{q}+i\eta} \delta_{qr} (\partial_{\nu}\mathcal{H})_{rs} \sigma_{sp}^{\mathrm{ss}}
+\mathrm{H.c.}\notag\\
=&i\frac{\hbar}{2}
\sum_{pqs}(\partial_{\mu}\mathcal{H})_{pq} \frac{1}{\epsilon_{p}-\epsilon_{q}+i\eta} (\partial_{\nu}\mathcal{H})_{qs} \frac{1}{\epsilon_{p}-\epsilon_{s}+i\eta} \Pi^{<}_{sp}
+\mathrm{H.c.}\notag\\
=&\frac{\hbar}{4\pi}
\int_{-\infty}^{\infty}d\epsilon\,\sum_{pqs}(\partial_{\mu}\mathcal{H})_{pq}\frac{1}{\epsilon-\epsilon_{q}+i\eta}(\partial_{\nu}\mathcal{H})_{qs}\frac{1}{\epsilon-\epsilon_{s}+i\eta}\Pi^{<}_{sp}\frac{1}{\epsilon-\epsilon_{p}-i\eta}
+\mathrm{H.c.}\notag\\
=&\frac{\hbar}{4\pi}
\int_{-\infty}^{\infty}d\epsilon\,\Tr{\partial_{\mu}\mathcal{H} \Ropt{\mathcal{G}}(\epsilon) \partial_{\nu}\mathcal{H} \mathcal{G}^{<}(\epsilon)}
+\mathrm{H.c.}\label{eq:D1}
\end{align}
Notice that the assumption that $\Pi^{<}$ is independent of $\epsilon$ is necessary for the equality from the second line to the third line.
Second we focus on the latter,
\begin{align}
&-i\frac{\hbar}{2}
\sum_{pqrs}(\partial_{\mu}\mathcal{H})_{pq} \frac{1}{\epsilon_{p}-\epsilon_{q}+i\eta} \sigma^{\mathrm{ss}}_{qr} (\partial_{\nu}\mathcal{H})_{rs} \sigma_{sp}^{\mathrm{ss}}
+\mathrm{H.c.}\notag\\
=&-i\frac{\hbar}{2}
\sum_{pqrs}(\partial_{\mu}\mathcal{H})_{pq} \frac{1}{\epsilon_{p}-\epsilon_{q}+i\eta}
\frac{1}{\epsilon_{r}-\epsilon_{q}+i\eta}\Pi^{<}_{qr}
(\partial_{\nu}\mathcal{H})_{rs}
\frac{1}{\epsilon_{p}-\epsilon_{s}+i\eta}\Pi^{<}_{sp}
+\mathrm{H.c.}\notag\\
=&-\frac{\hbar}{4\pi}
\int_{-\infty}^{\infty}d\epsilon\,
\sum_{pqrs}(\partial_{\mu}\mathcal{H})_{pq} \frac{1}{\epsilon-\epsilon_{q}+i\eta}
\frac{1}{\epsilon_{r}-\epsilon_{q}+i\eta}\Pi^{<}_{qr}
(\partial_{\nu}\mathcal{H})_{rs}
\frac{1}{\epsilon-\epsilon_{s}+i\eta}\Pi^{<}_{sp}\frac{1}{\epsilon-\epsilon_{p}-i\eta}
+\mathrm{H.c.}\notag\\
=&-\frac{\hbar}{4\pi}
\int_{-\infty}^{\infty}d\epsilon\,
\sum_{pqrs}(\partial_{\mu}\mathcal{H})_{pq} \frac{1}{\epsilon_{q}-\epsilon-i\eta}
\Pi^{<}_{qr}\frac{1}{\epsilon_{q}-\epsilon_{r}-i\eta}
(\partial_{\nu}\mathcal{H})_{rs}
\left(\mathcal{G}^{<}(\epsilon)\right)_{sp}
+\mathrm{H.c.}\notag\\
=&-i\frac{\hbar}{8\pi^{2}}
\int_{-\infty}^{\infty}d\epsilon\int_{-\infty}^{\infty}d\epsilon'\,
\sum_{pqrs}(\partial_{\mu}\mathcal{H})_{pq} \frac{1}{\epsilon'-\epsilon-i\eta}
\frac{1}{\epsilon'-\epsilon_{q}+i\eta}\Pi^{<}_{qr}\frac{1}{\epsilon'-\epsilon_{r}-i\eta}
(\partial_{\nu}\mathcal{H})_{rs}
\left(\mathcal{G}^{<}(\epsilon)\right)_{sp}
+\mathrm{H.c.}\notag\\
=&-i\frac{\hbar}{8\pi^{2}}
\int_{-\infty}^{\infty}d\epsilon\int_{-\infty}^{\infty}d\epsilon'\,
\frac{1}{\epsilon'-\epsilon-i\eta}
\Tr{\partial_{\mu}\mathcal{H} \mathcal{G}^{<}(\epsilon') \partial_{\nu}\mathcal{H} \mathcal{G}^{<}(\epsilon)}
+\mathrm{H.c.}\label{eq:D2}
\end{align}
Recall that only the ``symmetric'' part of $\bar{D}_{\mu\nu}^{\mathrm{S}}$ is meaningful in the Fokker-Planck equation, Eq. (\ref{eq:fokker_planck}). Therefore, it is proper to symmetrize Eq. (\ref{eq:D2}),
\begin{align}
&-\frac{i}{2}\frac{\hbar}{8\pi^{2}}\int_{-\infty}^{\infty}d\epsilon\int_{-\infty}^{\infty}d\epsilon'\,\left(\frac{1}{\epsilon'-\epsilon-i\eta}+\frac{1}{\epsilon-\epsilon'-i\eta}\right)\Tr{\partial_{\mu}\mathcal{H}\mathcal{G}^{<}(\epsilon')\partial_{\nu}\mathcal{H}\mathcal{G}^{<}(\epsilon)}+\mathrm{H.c.}\notag\\
=&\frac{\hbar}{8\pi}\int_{-\infty}^{\infty}d\epsilon\,\Tr{\partial_{\mu}\mathcal{H}\mathcal{G}^{<}(\epsilon)\partial_{\nu}\mathcal{G}^{<}(\epsilon)}+\mathrm{H.c.} \label{eq:sym_D2}
\end{align}
We must also symmetrize Eq. (\ref{eq:D1}). If we do so and add up both contributions, we obtain the final result:
\begin{align}
\frac{1}{2}\left(\bar{D}_{\mu\nu}^{\mathrm{S}}+\bar{D}_{\nu\mu}^{\mathrm{S}}\right)
=\frac{\hbar}{8\pi}\int_{-\infty}^{\infty}d\epsilon\bigg\{&
\Tr{\partial_{\mu}\mathcal{H}\Ropt{\mathcal{G}}(\epsilon)\partial_{\nu}\mathcal{H}\mathcal{G}^{<}(\epsilon)}
+\Tr{\partial_{\nu}\mathcal{H}\Ropt{\mathcal{G}}(\epsilon)\partial_{\mu}\mathcal{H}\mathcal{G}^{<}(\epsilon)}\notag\\
&+\Tr{\partial_{\mu}\mathcal{H}\mathcal{G}^{<}(\epsilon)\partial_{\nu}\mathcal{H}\mathcal{G}^{<}(\epsilon)}
\bigg\}+\mathrm{H.c.}\label{eq:sym_D}
\end{align}

\section{Calculation of the Symmetrized $\bar{D}_{\mu\nu}^{\mathrm{S}}$ with a Model Hamiltonian}\label{sec:D_222model}
In this section, we consider the molecular junction Hamiltonian (Eqs. (\ref{eq:total_e_H})-(\ref{eq:Hc})), and we simplify Eq. (\ref{eq:sym_D}). We further derive an analytic form of $\frac{1}{2}\left(\bar{D}_{\mu\nu}^{\mathrm{S}}+\bar{D}_{\nu\mu}^{\mathrm{S}}\right)$ when the two-orbital two-mode system Hamiltonian (namely $\mathbf{h}^{\mathrm{s}\uparrow}=\mathbf{h}(x,y)\cdot\bm{\sigma}$ mentioned in the main body of the text) is considered. As in Ref. \cite{teh2021antisymmetric}, if we consider the Condon limit, namely where  $V_{m,k\alpha}$ independent of $\mathbf{R}$, the trace in Eq. (\ref{eq:sym_D}) is taken over only the molecular orbitals. Therefore,
\begin{align}
\frac{1}{2}\left(\bar{D}_{\mu\nu}^{\mathrm{S}}+\bar{D}_{\nu\mu}^{\mathrm{S}}\right)
=\frac{\hbar}{8\pi}\int_{-\infty}^{\infty}d\epsilon\bigg\{&
\Tr{\partial_{\mu}h\Ropt{G}(\epsilon)\partial_{\nu}hG^{<}(\epsilon)}
+\Tr{\partial_{\nu}h\Ropt{G}(\epsilon)\partial_{\mu}hG^{<}(\epsilon)}\notag\\
&+\Tr{\partial_{\mu}hG^{<}(\epsilon)\partial_{\nu}hG^{<}(\epsilon)}
\bigg\}+\mathrm{H.c.},\label{eq:sym_D_condon}
\end{align}
where
\begin{align*}
\Ropt{G}(\epsilon)=\frac{1}{\epsilon-h-\Ropt{\Sigma}}
\end{align*}
is the molecule retarded Green's function with $\Ropt{\Sigma}$ denoting the molecule retarded self-energy,
\begin{align*}
\Ropt{\Sigma}_{mn}=\sum_{k\alpha}V_{m,k\alpha}\Ropt{g}_{k\alpha}V_{k\alpha,n},
\end{align*}
($\Ropt{g}_{k\alpha}=(\epsilon-\epsilon_{k\alpha}+i\eta)^{-1}$ is the lead retarded self-energy) and $G^{<}(\epsilon)$ is the molecule lesser Green's function.

Next, we specifically focus on the two-orbital two-mode model Hamiltonian,
which is a minimal model allowing us to see the nuclear Berry curvature effects. Under the standard wide-band-limit approximation, the tunneling-width matrix $\Gamma_{mn}\equiv2\pi\sum_{k\alpha}V_{m,k\alpha}V_{n,k\alpha}^{*}\delta(\epsilon-\epsilon_{k\alpha})$ is independent of $\epsilon$, and so $\Ropt{\mathbf{\Sigma}}=-i\mathbf{\Gamma}/2$. Since the left lead couples only to orbital $1$ and the right lead couples only to orbital $2$ (with the two coupling constants the same real value $\tilde{\Gamma}$), $\Ropt{\mathbf{\Sigma}}=-i\tilde{\Gamma}\mathbf{I}_{2\times2}/2$. According to previous calculations in Ref. \cite{teh2021antisymmetric},
\begin{align}
\Ropt{G}&=\frac{1}{\tilde{\epsilon}^{2}-h^{2}}\left(\tilde{\epsilon}+\mathbf{h}\cdot\bm{\sigma}\right),\label{eq:GR_2orb2mode}\\
G^{<}&=i\tilde{\Gamma}\abs{\frac{1}{\tilde{\epsilon}^{2}-h^{2}}}^{2}\left(\kappa_{0}+\bm{\kappa}\cdot\bm{\sigma}\right),\label{eq:Glesser_2orb2mode}
\end{align}
where $\tilde{\epsilon}$ and the $\kappa$'s are defined in Sec. \ref{sec:sum_friction_tensor}.

By using Eqs. (\ref{eq:GR_2orb2mode}) and (\ref{eq:Glesser_2orb2mode}), Eq. (\ref{eq:sym_D_condon}) becomes ($\hbar=1$)
\begin{align}
\frac{1}{2}\left(\bar{D}_{\mu\nu}^{\mathrm{S}}+\bar{D}_{\nu\mu}^{\mathrm{S}}\right)=\frac{1}{2\pi}\int_{-\infty}^{\infty}d\epsilon\,\bigg\{
&-2\mathfrak{Re}\{C'\}(\partial_{\mu}\mathbf{h}\cdot\partial_{\nu}\mathbf{h})(\mathbf{h}\cdot\bm{\kappa})\notag\\
&+2\mathfrak{Re}\{C'\}(\partial_{\mu}\mathbf{h}\cdot\mathbf{h})(\partial_{\nu}\mathbf{h}\cdot\bm{\kappa})\notag\\
&+2\mathfrak{Re}\{C'\}(\partial_{\nu}\mathbf{h}\cdot\mathbf{h})(\partial_{\mu}\mathbf{h}\cdot\bm{\kappa})\notag\\
&+2\kappa_{0}\mathfrak{Re}\{C'\tilde{\epsilon}\}\partial_{\mu}\mathbf{h}\cdot\partial_{\nu}\mathbf{h}\notag\\
&+C''\left[2(\partial_{\mu}\mathbf{h}\cdot\bm{\kappa})(\partial_{\nu}\mathbf{h}\cdot\bm{\kappa})+(\kappa_{0}^{2}-\kappa^{2})\partial_{\mu}\mathbf{h}\cdot\partial_{\nu}\mathbf{h}\right]
\bigg\},\label{eq:sym_D_practical}
\end{align}
where
\begin{align*}
C'\equiv&\left(\frac{1}{\tilde{\epsilon}^{2}-h^{2}}\right)i\tilde{\Gamma}\abs{\frac{1}{\tilde{\epsilon}^{2}-h^{2}}}^{2},\\
C''\equiv&-\tilde{\Gamma}^{2}\abs{\frac{1}{\tilde{\epsilon}^{2}-h^{2}}}^{4}.
\end{align*}
Equation (\ref{eq:sym_D_practical}) is the covariance matrix we evaluate in practice for propagating the Langevin equation, Eq. (\ref{eq:langevin_eq}).



\section{Positive Definiteness of $(\bar{D}_{\mu\nu}^{\mathrm{S}}+\bar{D}_{\nu\mu}^{\mathrm{S}})/2$}\label{sec:positive_definiteness_D}
In this section, we prove that the symmetrized covariance matrix $(\bar{D}_{\mu\nu}^{\mathrm{S}}+\bar{D}_{\nu\mu}^{\mathrm{S}})/2$ is positive definite for a complex-valued Hamiltonian when the system is in/out of equilibrium. This property enables us to utilize the Cholesky decomposition to sample the random force. We start by noticing that, since $\left[\hat{H},\hat{\rho}_{\mathrm{ss}}\right]=0$, we can always choose a unique Lehmann representation as an eigenbasis for both $\hat{H}$ and $\hat{\rho}_{\mathrm{ss}}$, namely $\hat{H}\vert m\rangle=E_{m}\vert m\rangle$ and $\hat{\rho}_{\mathrm{ss}}\vert m\rangle=\rho_{m}\vert m\rangle$. (Note that $\rho_{m}>0$ because a density matrix is positive definite.) Under this representation, the general expression for the covariance matrix $\bar{D}_{\mu\nu}^{\mathrm{S}}$ in Eq. (\ref{eq:D}) becomes
\begin{align*}
\bar{D}_{\mu\nu}^{\mathrm{S}}=\frac{i\hbar}{2}\sum_{mn}\frac{1}{E_{n}-E_{m}+i\eta}\langle n\vert\delta\hat{F}_{\mu}\vert m\rangle\langle m\vert\delta\hat{F}_{\nu}\vert n\rangle\left(\rho_{m}+\rho_{n}\right),
\end{align*}
where we have used integral representations of the Dirac delta function and the Heaviside function. We then symmetrize the covariance matrix,
\begin{align*}
\frac{1}{2}(\bar{D}_{\mu\nu}^{\mathrm{S}}+\bar{D}_{\nu\mu}^{\mathrm{S}})
&=\frac{i\hbar}{2}\sum_{mn}\frac{1}{E_{n}-E_{m}+i\eta}
\left(\langle n\vert\delta\hat{F}_{\mu}\vert m\rangle\langle m\vert\delta\hat{F}_{\nu}\vert n\rangle
+\langle n\vert\delta\hat{F}_{\nu}\vert m\rangle\langle m\vert\delta\hat{F}_{\mu}\vert n\rangle\right)
\left(\rho_{m}+\rho_{n}\right)\\
&=\frac{i\hbar}{2}\sum_{mn}\left(\frac{1}{E_{n}-E_{m}+i\eta}+\frac{1}{E_{m}-E_{n}+i\eta}\right)\langle n\vert\delta\hat{F}_{\mu}\vert m\rangle\langle m\vert\delta\hat{F}_{\nu}\vert n\rangle\left(\rho_{m}+\rho_{n}\right)\\
&=\pi\hbar\sum_{mn}\delta(E_{n}-E_{m})\langle n\vert\delta\hat{F}_{\mu}\vert m\rangle\langle m\vert\delta\hat{F}_{\nu}\vert n\rangle\left(\rho_{m}+\rho_{n}\right),
\end{align*}
where we have used the representation of the Dirac delta function, $\lim_{\epsilon\rightarrow0}\epsilon/\pi(x^{2}+\epsilon^{2})=\delta(x)$. Thus, for arbitrary real vectors $\mathbf{X}\neq0$, we have
\begin{align*}
\sum_{\mu\nu}X_{\mu}\frac{1}{2}(\bar{D}_{\mu\nu}^{\mathrm{S}}+\bar{D}_{\nu\mu}^{\mathrm{S}})X_{\nu}
=\pi\hbar\sum_{mn}\delta(E_{n}-E_{m})(\rho_{m}+\rho_{n})\abs{\langle n\vert\left(\sum_{\mu}X_{\mu}\delta\hat{F}_{\mu}\right)\vert m\rangle}^{2}>0.
\end{align*}
Hence we have proven that $(\bar{D}_{\mu\nu}^{\mathrm{S}}+\bar{D}_{\nu\mu}^{\mathrm{S}})/2$ is always positive definite.

\section{Fluctuation-Dissipation Theorem (Non-interacting Hamiltonian)}\label{sec:fluctuation_dissipation}
In this section, we will prove that at equilibrium the fluctuation-dissipation theorem is still obeyed between $\gamma_{\mu\nu}^{\mathrm{S}}$ and $\left(\bar{D}_{\mu\nu}^{\mathrm{S}}+\bar{D}_{\nu\mu}^{\mathrm{S}}\right)/2$ which is derived in SM \ref{sec:D_in_terms_of_G}. Equation (\ref{eq:sym_D}) can be recast into a simpler form by using the relation $\Ropt{\mathcal{G}}-\Aopt{\mathcal{G}}=\mathcal{G}^{>}-\mathcal{G}^{<}$,
\begin{align*}
\frac{1}{2}\left(\bar{D}_{\mu\nu}^{\mathrm{S}}+\bar{D}_{\nu\mu}^{\mathrm{S}}\right)
=\frac{\hbar}{8\pi}\int_{-\infty}^{\infty}d\epsilon\,\Tr{\partial_{\mu}\mathcal{H}\mathcal{G}^{>}\partial_{\nu}\mathcal{H}\mathcal{G}^{<}}+\mathrm{H.c.},
\end{align*}
which can be further simplified when $\mathcal{G}^{<}$ is anti-Hermitian,
\begin{align}
\frac{1}{2}\left(\bar{D}_{\mu\nu}^{\mathrm{S}}+\bar{D}_{\nu\mu}^{\mathrm{S}}\right)
=\frac{\hbar}{4\pi}\int_{-\infty}^{\infty}d\epsilon\,\Tr{\partial_{\mu}\mathcal{H}\mathcal{G}^{>}\partial_{\nu}\mathcal{H}\mathcal{G}^{<}}.
\end{align}
Note that in equilibrium $\mathcal{G}^{<}$ is anti-Hermitian because $\mathcal{G}^{<}=-f(\Ropt{\mathcal{G}}-\Aopt{\mathcal{G}})$.
Next, we symmetrize Eq. (\ref{eq:gamma_noninteracting_antiH_glesser}) and consider the equilibrium situation,
\begin{align}
\gamma_{\mu\nu}^{\mathrm{S}}
=&\frac{\hbar}{4\pi}\int_{-\infty}^{\infty}d\epsilon\,
\Tr{\partial_{\mu}\mathcal{H}\partial_{\epsilon}\Ropt{\mathcal{G}}\partial_{\nu}\mathcal{H}\mathcal{G}^{<}}
+(\mu\leftrightarrow\nu)
+\mathrm{H.c.}\notag\\
=&-\frac{\hbar}{4\pi}\int_{-\infty}^{\infty}d\epsilon\,
\Tr{\partial_{\mu}\mathcal{H}\Ropt{\mathcal{G}}\partial_{\nu}\mathcal{H}\partial_{\epsilon}\mathcal{G}^{<}}
+(\mu\leftrightarrow\nu)
+\mathrm{H.c.}\notag\\
=&\frac{\hbar}{4\pi}\int_{-\infty}^{\infty}d\epsilon\,
\bigg\{
\Tr{\partial_{\mu}\mathcal{H}\Ropt{\mathcal{G}}\partial_{\nu}\mathcal{H}(\Ropt{\mathcal{G}}-\Aopt{\mathcal{G}})}\partial_{\epsilon}f
+\Tr{\partial_{\mu}\mathcal{H}\Ropt{\mathcal{G}}\partial_{\nu}\mathcal{H}\partial_{\epsilon}(\Ropt{\mathcal{G}}-\Aopt{\mathcal{G}})}f
\bigg\}\notag\\
&+(\mu\leftrightarrow\nu)
+\mathrm{H.c.}\notag\\
=&\frac{\hbar}{4\pi}\int_{-\infty}^{\infty}d\epsilon\,
\Tr{\partial_{\mu}\mathcal{H}(\Ropt{\mathcal{G}}-\Aopt{\mathcal{G}})\partial_{\nu}\mathcal{H}(\Ropt{\mathcal{G}}-\Aopt{\mathcal{G}})}\partial_{\epsilon}f\notag\\
=&\frac{\beta\hbar}{4\pi}\int_{-\infty}^{\infty}d\epsilon\,\Tr{\partial_{\mu}\mathcal{H}\mathcal{G}^{>}\partial_{\nu}\mathcal{H}\mathcal{G}^{<}}\notag\\
=&\beta\frac{1}{2}\left(\bar{D}_{\mu\nu}^{\mathrm{S}}+\bar{D}_{\nu\mu}^{\mathrm{S}}\right).\label{eq:fluctuation_dissipation}
\end{align}
Thus, the fluctuation-dissipation theorem is satisfied at equilibrium.

With the aid of Eq. (\ref{eq:fluctuation_dissipation}), one can determine the steady state density distribution 
using only $F_{\mu}$
(and without knowledge of $\gamma_{\mu \nu}$ or $\bar{D}_{\mu\nu}^{\mathrm{S}}$). To prove this fact, we need only show that the simplest Boltzmann distribution guess for $\rho$,
\begin{align}
\label{guess}
\rho=\frac{e^{-\beta\left[V(\mathbf{R})+\sum_{\alpha}P_{\alpha}^{2}/2m_{\alpha}\right]}}{Z},
\end{align}
satisfies $\partial_t \rho =0$ (see Eq. (\ref{eq:fokker_planck})).
Here $Z$ is the partition function.
We simply plug Eq. (\ref{guess}) into the right hand side of Eq. (\ref{eq:fokker_planck}).
The third term becomes:
\begin{align*}
\sum_{\mu\nu}\gamma_{\mu\nu}\frac{\partial}{\partial P_{\mu}}\left(\frac{P_{\nu}}{m_{\nu}}\rho\right)
=&\sum_{\mu\nu}\gamma_{\mu\nu}^{\mathrm{S}}\frac{\partial}{\partial P_{\mu}}\left(\frac{P_{\nu}}{m_{\nu}}\rho\right)+
\sum_{\mu\nu}\gamma_{\mu\nu}^{\mathrm{A}}
\left[\frac{\rho}{m_{\nu}}\delta_{\mu\nu}-\beta\rho\frac{P_{\mu}P_{\nu}}{m_{\mu}m_{\nu}}\right]\\
=&\sum_{\mu\nu}\gamma_{\mu\nu}^{\mathrm{S}}\frac{\partial}{\partial P_{\mu}}\left(\frac{P_{\nu}}{m_{\nu}}\rho\right).
\end{align*}
According to the fluctuation-dissipation theorem in Eq. (\ref{eq:fluctuation_dissipation}), the fourth term on the RHS of Eq. (\ref{eq:fokker_planck}) becomes
\begin{align*}
\sum_{\mu\nu}\frac{1}{2}\left(\bar{D}_{\mu\nu}^{\mathrm{S}}+\bar{D}_{\nu\mu}^{\mathrm{S}}\right)\frac{\partial^{2}\rho}{\partial P_{\mu}\partial P_{\nu}}=
\sum_{\mu\nu}\frac{\gamma_{\mu\nu}^{\mathrm{S}}}{\beta}
\frac{\partial}{\partial P_{\mu}}\left(-\beta\rho\frac{P_{\nu}}{m_{\nu}}\right),
\end{align*}
which cancels with the third term. (Recall that, as mentioned in Sec. \ref{sec:D_in_terms_of_G}, the antisymmetric part of $\bar{D}_{\mu\nu}^{\mathrm{S}}$ does not enter the Fokker-Planck equation because $\partial^{2}\rho/\partial P_{\mu}\partial P_{\nu}$ is symmetric.) Also, the first and the second terms on the RHS cancel with each other,
\begin{align*}
-\sum_{\mu}\frac{P_{\mu}}{m_{\mu}}\partial_{\mu}\rho-\sum_{\mu}F_{\mu}\frac{\partial\rho}{\partial P_{\mu}}=-\sum_{\mu}\frac{P_{\mu}}{m_{\mu}}\rho(-\beta)\partial_{\mu}V-\sum_{\mu}\left(-\partial_{\mu}V\right)\rho(-\beta)\frac{P_{\mu}}{m_{\mu}}=0.
\end{align*}
Thus, the Boltzmann distribution is a steady state solution ($\partial_{t}\rho=0$). In other words, at equilibrium (where fluctuation dissipation holds), one can use $F_{\mu}$ alone to obtain the steady state probability distribution. However, out of equilibrium, there is no such guarantee and, in the main body of the paper, we show that when there is a current present, $\rho$ can depend critically on $\gamma_{\mu\nu}$ (and be very different for $\left(\gamma_{\mu\nu}^{\mathrm{A}}\right)^{\uparrow/\downarrow}$).

\section{Adiabatic Force}\label{sec:adiaF}
In order to run Langevin dynamics, one requires the friction tensor, a random force and the adiabatic force. So far, we have treated the first two quantities, and what remains is to calculate the adiabatic force. As has been discussed at great length, such a force is non-conservative out of equilibrium in the presence of a current. In order to calculate such the force in Eq. (\ref{eq:adia_F}),
we plug Eq. (\ref{eq:glesser_n_sigma}) into Eq. (\ref{eq:adia_F}),
\begin{align*}
F_{\mu}
=&-\sum_{pq}\mathcal{H}_{pq}(\mathbf{R})\sigma_{qp}^{\mathrm{ss}}-\partial_{\mu}U_{0}(\mathbf{R})\\
=&-\frac{1}{2\pi i}\int_{-\infty}^{\infty}d\epsilon\,\Tr{\partial_{\mu}\mathcal{H}\mathcal{G}^{<}}-\partial_{\mu}U_{0}
\end{align*}
We further make the Condon approximation such that only the system Hamiltonian (rather than the system-bath Hamiltonian) changes as a function of nuclear coordinate:
\begin{align*}
F_{\mu}=-\frac{1}{2\pi i}\int_{-\infty}^{\infty}d\epsilon\,\Tr{\partial_{\mu}hG^{<}}-\partial_{\mu}U_{0}.
\end{align*}
Finally, in our two-orbital model, according to Eq. (\ref{eq:Glesser_2orb2mode}),  the adiabatic force becomes
\begin{align*}
F_{\mu}=-\frac{\tilde{\Gamma}}{\pi}\int_{-\infty}^{\infty}\abs{\frac{1}{\tilde{\epsilon}^{2}-h^{2}}}^{2}\partial_{\mu}\mathbf{h}\cdot\bm{\kappa}-\partial_{\mu}U_{0}.
\end{align*}

\section{Transmission Probability}\label{sec:transmission_prob}
The transmission probability in the Landauer formula, Eq. (\ref{eq:Landauer_formula_local_current}), can be expressed in terms of Green's functions (see Ref. \cite{haug2008quantum} or \cite{nitzan2006chemical} for details):
\begin{align}
T(\epsilon)=\Tr{\Gamma^{\mathrm{L}} \Ropt{G}(\epsilon) \Gamma^{\mathrm{R}} \Aopt{G}(\epsilon)}.\label{eq:transmission_prob_no_superscript}
\end{align}
Within our setup, only orbital $1$  couples to the left lead and only orbital $2$ couples to the right lead. Thus, the $\Gamma$ matrices are
\begin{align*}
\Gamma^{\mathrm{L}}=
\begin{pmatrix}
\tilde{\Gamma} & 0\\
0 & 0
\end{pmatrix},
\quad
\Gamma^{\mathrm{R}}=
\begin{pmatrix}
0 & 0\\
0 & \tilde{\Gamma}
\end{pmatrix}.
\end{align*}
Hence,
\begin{align*}
T(\epsilon)=\tilde{\Gamma}^{2}\Ropt{G}_{12}\Aopt{G}_{21}=\tilde{\Gamma}^{2}\left(h_{1}^{2}+h_{2}^{2}\right)\abs{\frac{1}{\tilde{\epsilon}^{2}-h^{2}}}^{2}.
\end{align*}
Note that  $T(\epsilon)$ is invariant to changing
$h_{2}\rightarrow-h_{2}$, which implies that the local current $I_{\mathrm{loc}}^{\uparrow/\downarrow}$ is in fact independent of the exact spin carrier.
For this reason, we have not included any superscripts $\uparrow/\downarrow$ in Eq. (\ref{eq:transmission_prob_no_superscript}).

\section{Enhancement of the Spin Polarization with Nonzero $\Delta$}\label{sec:D3}
\begin{figure}[!h]
\centering
\includegraphics[width=.6\textwidth]{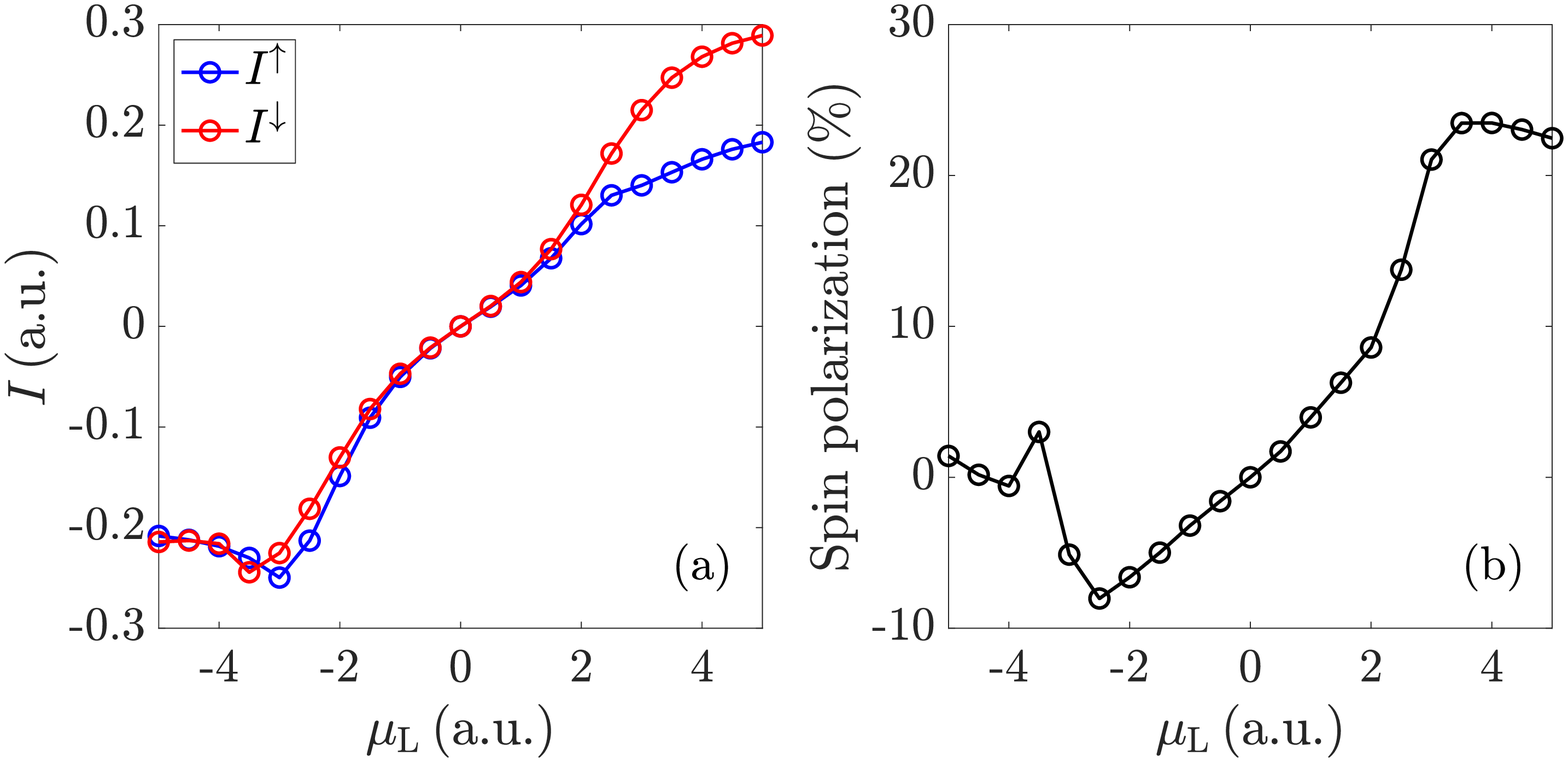}
\caption{Calculations for (a) spin currents and (b) the corresponding spin polarization when the voltage bias is nonzero. Parameters: $\Delta=3$, $A=B=1$, $\kappa_{x}=0$, $\chi=1$ and $\kappa=1$. The spin polarization can be enhanced when $\Delta\neq0$.\label{fig:D3}}
\end{figure}

In Fig. \ref{fig:D3} (a) and (b), we utilize Eq. (\ref{eq:spin_current}) in the main body of the text to calculate the spin currents and the corresponding spin polarization in the presence of a nonzero energy gap $\Delta=3$ between the two orbitals. Compared to the case $\Delta=0$, the spin polarization is enhanced for positive $\mu_{\mathrm{L}}$ but is diminished  for negative $\mu_{\mathrm{L}}$.

\section{Spin Current and Spin Polarization for A Small Spin-Orbit Coupling}\label{sec:small_soc}
\begin{figure}[!h]
\centering
\includegraphics[width=.6\textwidth]{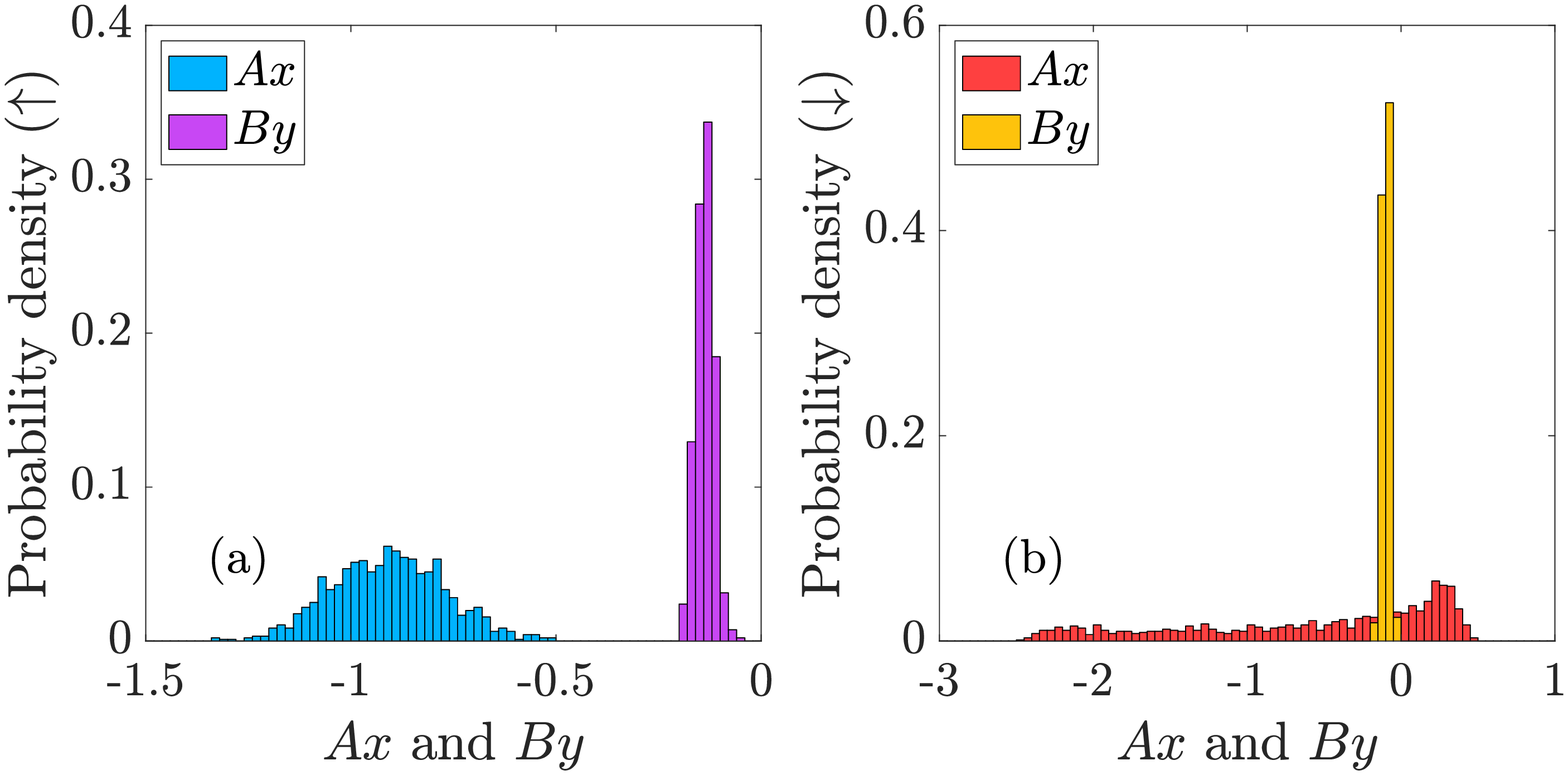}
\caption{Probability distribution of $Ax$ and $By$ for (a) spin up (b) spin down carriers.\label{fig:Ax_By}}
\end{figure}

\begin{figure}[!h]
\centering
\includegraphics[width=.6\textwidth]{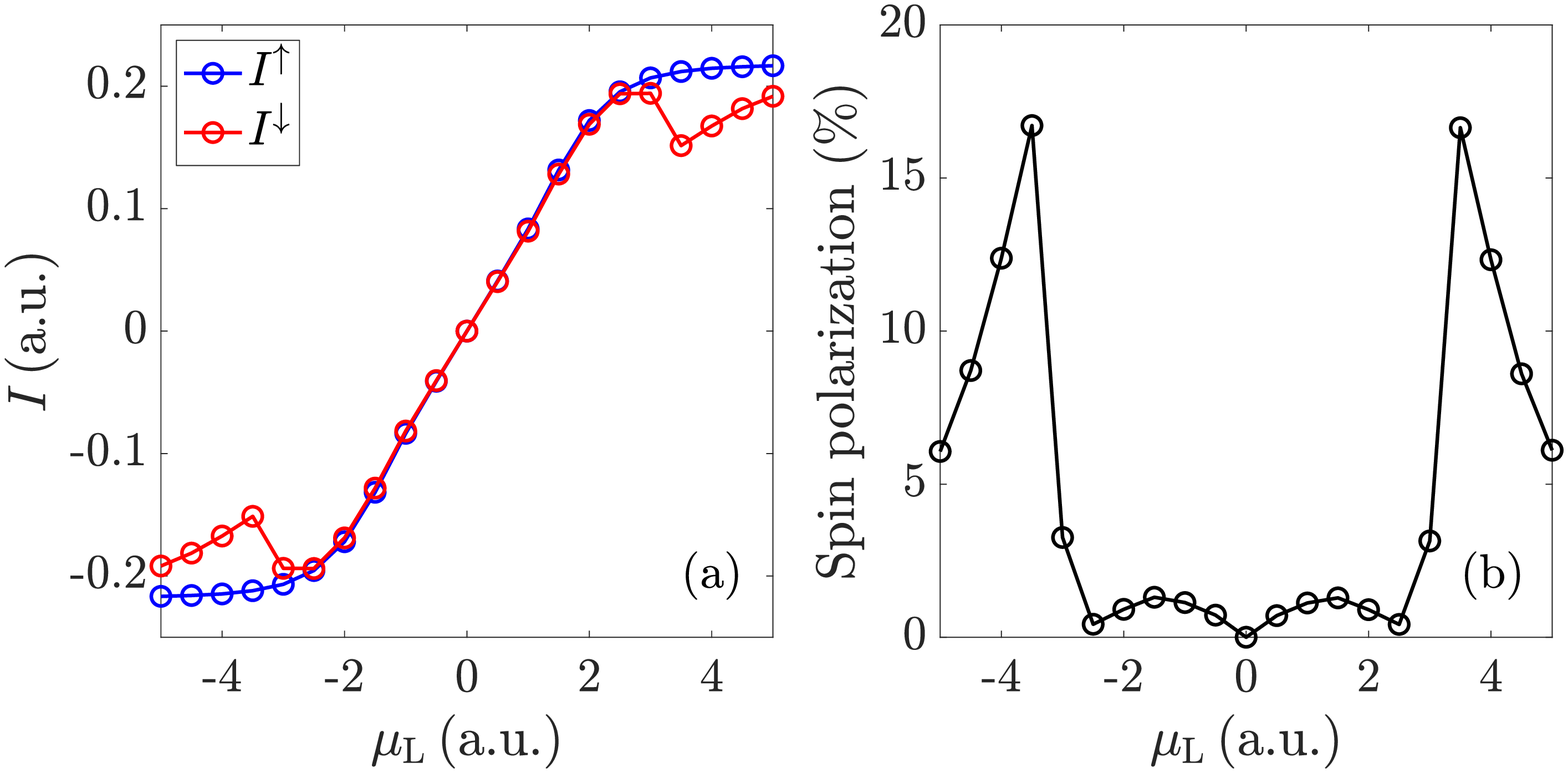}
\caption{Calculations for (a) spin currents and (b) the corresponding spin polarization when the spin-orbit coupling is small. Parameters: $\Delta=0$, $A=1$, $B=0.1$, $\kappa_{x}=0$, $\chi=0.1$ and $\kappa=0.1$. Sizable spin polarization can still be achieved when different mode frequencies are considered.\label{fig:small_soc}}
\end{figure}

We have not yet formally addressed the question of the size of the spin-orbit interaction. One can ask: can reasonable spin polarization emerge if the spin-orbit interaction is not too large?  To answer such a question,
in Figs. \ref{fig:small_soc} (a) and (b), we calculate the spin currents and the corresponding spin polarization with a smaller spin-orbit interaction.
More specifically, we reduce both $B$ (in Eq. (\ref{eq:hs_spin_up})) and $\chi$ (in Eq. (\ref{eq:U})): we set $B = \chi = 0.1$. While reducing $\chi$ should lead to larger fluctuations in the position $y$, reducing $B$ leads to a smaller total spin-orbit coupling matrix element.
In Fig. \ref{fig:Ax_By}, we show  a histogram of the resulting diabatic couplings ($Ax$) and spin-orbit couplings ($By$); note that indeed we have reduced the total size of the average spin-orbit coupling relative to the average diabatic coupling. In Fig. \ref{fig:small_soc}, we then show the resulting currents and spin-polarization. Observe that a meaningful spin-polarization can indeed be obtained, even when the spin-orbit coupling matrix elements are one tenth the size of the diabatic coupling matrix elements.  


\section{Shifted Parabola Model}\label{sec:shifted_parabola}
In this section, we illustrate our two-orbital system Hamiltonian used in the main body of text in details. The combination of Eqs. (\ref{eq:hs_spin_up}) and (\ref{eq:U}) is a very general form of model Hamiltonian describing two shifted parabola in the nuclear space. The parameters $A$ and $B$ describe how fast the diabatic coupling changes as the geometry ($x$ and $y$) of the molecule changes. $\Delta$ controls the energy gap between two parabola, and $\chi$ tunes the ratio between frequencies of the two modes. This shifted parabola model is commonly used in simulating electron transfer as well as excitation energy transfer processes\cite{nitzan2006chemical}, where the initial (single particle) state is localized on one orbital with one nuclear distribution (geometry), and the final state localizes on the other orbital with another nuclear distribution. 

The parameters used in this letter correspond roughly to the \textit{ab initio} parameters extracted for a diphenylmethane junction where we considered the LUMO and LUMO+1 (Sec. J of the SM Ref. \cite{teh2021antisymmetric}). In particular, there we used linear functions $\lambda x +\Delta$ and $Ax+C$ to fit the site energy and real part of the diabatic coupling of the \textit{ab initio} data respectively; we extracted the parameters $\lambda=3.44\times10^{-4}$, $\Delta=-1.13\times10^{-4}$, $A=3.44\times10^{-4}$ and $C=1.11\times10^{-3}$ (all in atomic units). 
The coupling constant $\tilde{\Gamma}$ is chosen to range over standard values found in the literature ($10-100\,\mathrm{meV}$)\cite{guo2012spin,koseki2019spin,varela2016effective}.  By renormalizing all of the energies above with $\lambda$, one finds parameters that are consistent with the parameters used in this letter. The only variable that was not extracted in an \textit{ab initio} fashion is the spin-orbit coupling, which was difficult to assess from a small cluster calculation. Thus, above, we have explored the parameter region  $B=0.1-1A$ so that we can assess the form of dynamics as $B$ gets smaller.

\section{Block Diagonalization of Two-Orbital Two-Spin Hamiltonians}\label{sec:block_diag}
In this section, we derive Eq. (\ref{eq:hs_block_diag}) in more detail. We first consider the following general model Hamiltonian with spin-orbit interaction,
\begin{align*}
H=H_{0}+H_{\mathrm{SOC}},
\end{align*}
where $H_{0}$  is a function of only orbital degrees of freedom (DoF), and $H_{\mathrm{SOC}}=\xi\mathbf{L}\cdot\mathbf{S}$ captures spin-orbit coupling with coupling strength $\xi$. We will focus on a two-orbital two-spin model system, and our goal is to block diagonalize this Hamiltonian, decoupling spin DoF.

Written in the basis $\lbrace\vert1\uparrow\rangle,\vert1\downarrow\rangle,\vert2\uparrow\rangle,\vert2\downarrow\rangle\rbrace$, the most general $H_{0}$ is
\begin{align}
\mathbf{H}_{0}=
\begin{pmatrix}
E_{1} & 0 & V & 0\\
0 & E_{1} & 0 & V\\
V & 0 & E_{2} & 0\\
0 & V & 0 & E_{2}
\end{pmatrix},\label{eq:H0_before_reordering}
\end{align}
where $E_{1}$ and $E_{2}$ label orbital energies and $V$ denotes coupling between the two orbitals. The spin-orbit coupling matrix $\mathbf{H}_{\mathrm{SOC}}$ can be constructed by calculating matrix elements $\langle\alpha m\vert H_{\mathrm{SOC}}\vert\beta n\rangle=\xi\frac{\hbar}{2}\mathbf{L}_{mn}\cdot\langle\alpha\vert\bm{\sigma}\vert\beta\rangle$ where $m$ and $n$ label orbital $1$ and $2$, $\alpha$ and $\beta$ represent spin up and down electrons. Since the spatial orbitals $m$ and $n$ can always chosen to be real functions, $\mathbf{L}_{mn}$ is purely imaginary, and so $\mathbf{L}_{mm}=0$. Therefore,
\begin{align*}
\mathbf{H}_{\mathrm{SOC}}=
\xi\frac{\hbar}{2}
\begin{pmatrix}
0 & 0 & L_{12}^{z} & L_{12}^{x}-iL_{12}^{y}\\
0 & 0 & L_{12}^{x}+iL_{12}^{y} & -L_{12}^{z}\\
L_{21}^{z} & L_{21}^{x}-iL_{21}^{y} & 0 & 0\\
L_{21}^{x}+iL_{21}^{y} & -L_{21}^{z} & 0 & 0
\end{pmatrix}
\equiv
\begin{pmatrix}
\mathbf{0} & \mathbf{A}\\
\mathbf{A}^{\dagger} & \mathbf{0}
\end{pmatrix},
\end{align*}
where $\mathbf{A}$ is anti-Hermitian and can be diagonalized $\mathbf{A}=\mathbf{U}\mathbf{a}\mathbf{U}^{\dagger}$. Here $\mathbf{U}\mathbf{U}^{\dagger}=\mathbf{I}$ and
\begin{align*}
\mathbf{a}=
\begin{pmatrix}
a & 0\\
0 & -a
\end{pmatrix},
\end{align*}
where $a=i\xi\hbar\abs{\mathbf{L}_{12}}/2$ is purely imaginary (and so $a^{*}=-a$). We can then transform $\mathbf{H}_{\mathrm{SOC}}$ to a new basis $\lbrace\left|1\uparrow'\right>$,$\left|1\downarrow'\right>$,$\left|2\uparrow'\right>$,$\vert2\downarrow'\rangle\rbrace$ as follows,
\begin{align}
\mathbf{H}_{\mathrm{SOC}}\rightarrow
\mathbf{H}_{\mathrm{SOC}}'=
\begin{pmatrix}
\mathbf{U} & \mathbf{0}\\
\mathbf{0} & \mathbf{U}
\end{pmatrix}
\begin{pmatrix}
\mathbf{0} & \mathbf{A}\\
\mathbf{A}^{\dagger} & \mathbf{0}
\end{pmatrix}
\begin{pmatrix}
\mathbf{U}^{\dagger} & \mathbf{0}\\
\mathbf{0} & \mathbf{U}^{\dagger}
\end{pmatrix}
=
\begin{pmatrix}
\mathbf{0} & \mathbf{a}\\
\mathbf{a}^{\dagger} & \mathbf{0}
\end{pmatrix}.\label{eq:HSOC_before_reordering}
\end{align}
Note that this transformation involves a rotation only for spin DoF. That is, $\mathbf{H}_{0}$ is invariant under this transformation. By reordering the new basis $\lbrace\vert1\uparrow'\rangle,\vert1\downarrow'\rangle,\vert2\uparrow'\rangle,\vert2\downarrow'\rangle\rbrace$ to $\lbrace\vert1\uparrow'\rangle,\vert2\uparrow'\rangle,\vert1\downarrow'\rangle,\vert2\downarrow'\rangle\rbrace$ in Eqs. (\ref{eq:H0_before_reordering}) and (\ref{eq:HSOC_before_reordering}), we recover Eq. (\ref{eq:hs_block_diag}).


\end{document}